\documentclass{article}

\usepackage[final]{neurips_2022}

\usepackage[utf8]{inputenc} 
\usepackage[T1]{fontenc}    
\usepackage{hyperref}       
\usepackage{url}            
\usepackage{booktabs}       
\usepackage{amsfonts}       
\usepackage{microtype}      
\usepackage{float}
\usepackage[table,xcdraw]{xcolor}
\usepackage[ruled,vlined]{algorithm2e}
\usepackage{setspace}
   
\usepackage{fancyhdr}

 \usepackage{tikz}
 \usetikzlibrary{arrows}

\title{Detecting Abrupt Changes in  Sequential Pairwise Comparison Data}

\author{%
  Wanshan Li\\
  Department of Statistics \& Data Science\\
  Carnegie Mellon University\\
  \texttt{wanshanl@andrew.cmu.edu} 
   \And
   Daren Wang \\
   Department of ACMS\\
University of Notre Dame \\
   \texttt{dwang24@nd.edu}
\And
   Alessandro Rinaldo \\
    Department of Statistics \& Data Science\\
  Carnegie Mellon University\\
   \texttt{arinaldo@cmu.edu}
}

\usepackage{./Latex/macros_01_proof_envs}
\usepackage{./Latex/macros_02_operators_commands}
\usepackage{./Latex/macros_03_symbols}
\usepackage{./Latex/macros_04_propernames}
\usepackage{./Latex/macros_05_reviewer}
\allowdisplaybreaks

\newcommand{\btl}{BTL}

\usepackage{csquotes}

\usepackage{array}
\newcommand{\PreserveBackslash}[1]{\let\temp=\\#1\let\\=\temp}
\newcolumntype{C}[1]{>{\PreserveBackslash\centering}p{#1}}
\newcolumntype{R}[1]{>{\PreserveBackslash\raggedleft}p{#1}}
\newcolumntype{L}[1]{>{\PreserveBackslash\raggedright}p{#1}}

\begin{document}

\maketitle

\begin{abstract}
    The Bradley-Terry-Luce (\btl{}) model is a classic and very popular statistical approach  for eliciting a global ranking among a collection of items using pairwise comparison data. In applications in which the comparison outcomes  are observed as a time series, it is often the case that data are non-stationary, in the sense that the true underlying ranking changes over time. In this paper we are concerned with localizing the change points in a high-dimensional \btl{} model with piece-wise constant parameters. We propose novel and practicable algorithms based on dynamic programming that can consistently estimate the unknown locations of the change points. We provide consistency rates for our methodology that depend explicitly on the model parameters, the temporal spacing between two consecutive change points and the magnitude of the change. We corroborate our findings with extensive numerical experiments and a real-life example.
\end{abstract}

\section{Introduction}
\label{sec:introduction}

Pairwise comparison data are among the most common types of data collected for the purpose of eliciting a global ranking among a collection of items or teams. The Bradley-Terry-Luce model \citep{bradley1952rank,Luce1959} is a classical and popular parametric approach to model pairwise comparison data and to obtain an estimate of the underlying ranking. The Bradley-Terry-Luce model and its variants have been proven to be powerful approaches in many applications, including sports analytics \citep{FaT1994,MaV2012, CMV2012}, bibliometrics \citep{St1994, Va2016}, search analytics \citep{RaJ2007,Ag2002}, and much more.

To introduce the  \btl{} model, suppose that we are interested in ranking $n$
distinct items, each with a (fixed but unobserved) positive 
preference score $w_{i}$, $i \in [n]$, quantifying its propensity to beat other items in a pairwise comparison. The \btl{} model assumes that the outcomes of the comparisons between different pairs are  independent Bernoulli random variables such that, for a given pair of items, say  $i$ and  $j$ in $[n] := \{1,\ldots,n\}$, the probability that $i$ is preferred to (or beats)  $j$ is
\begin{equation}\label{neqn:bradley_terry_prob_succ}
P_{ij}
 = \Prb{i \text{ beats } j}
 = \frac{w^*_{i}}{w^*_{i} + w^*_j}, \: \forall \; i,j \in [n].
\end{equation}
A common reparametrization is to set $w^*_{i} =
\exp(\theta^*_i)$ for each $i$, where $\boldsymbol{\theta}^* \defined (\theta^*_{1}, \ldots,
\theta^*_{n})^{\top} \in \reals^{n}$. To ensure identifiability it is further assumed that $\sum_{i \in [n]} \theta^*_i = 0$.

The properties and performance of the \btl{} model have been thoroughly studied under the assumption that the outcomes of all the pairwise comparisons are simultaneously available and follow the same \btl{} model. 
In many applications however, it is very common to observe pairwise comparison data sequentially (i.e. one at a time), with time stamps over multiple time periods. In these cases, it is unrealistic to assume that observations with different time stamps come from the same distribution. For instance, in sports analytics, the performance of teams often changes across match rounds, and \cite{FaT1994} utilized a state-space generalization of the \btl{} model to analyze sport tournaments data. Ranking analysis with temporal variants has also become increasingly important because of the growing needs for models and methods to handle time-dependent data. A series of results in this direction can be found in \cite{Gli1993}, \cite{glickman1998statespace_nfl}, \cite{CMV2012}, \cite{lopez2018often}, \cite{MKG2019}, \cite{bong2020}, \cite{Karle2021} and references therein.
Much of the aforementioned literature on time-varying \btl{} model postulates that temporal changes in the model parameters are smooth functions of time and thus occur gradually on a relatively large time scale. However, there are instances in which it may be desirable to instead model abrupt changes in the underlying parameters and estimate the times at which such change has occurred. These change point settings, which, to the best of our knowledge,  have not been considered in the literature, and are the focus of this paper. 


\
\\
{\bf Contributions}

We make the following methodological and theoretical contributions.

   $\bullet $ {\bf Novel change point methodology.}  
We develop a computationally efficient methodology to consistently estimate the change points for a time-varying \btl{} model with piece-wise constant parameters. Our baseline procedure \Cref{algo:DP} consists of a penalized maximum likelihood estimator of the \btl{} model under an $\ell_0$ penalty, and can be efficiently implemented via dynamic programming. We further propose a slightly more computationally expensive two-step procedure in \Cref{algo:local_refine_bt} that takes as input the estimator returned by our baseline procedure and delivers a more precise estimator with provably better error rates. We demonstrate through simulations and a real life example the performance and practicality of the procedure we develop.

 $\bullet $ {\bf Theoretical guarantees.}  We obtain finite sample error rates for our procedures that depend explicitly on all the parameters at play: the dynamic range of the \btl{} model and the number of items to be compared, the number of change points, the smallest distance between two consecutive change points and the minimal magnitude of the difference between the model parameters at two consecutive change points. Importantly, our theory allows for general connected \textit{comparison graph} and it explicitly captures the effect the topology of the comparison graph. Our results hold provided that a critical signal-to-noise ratio condition involving all the relevant parameters is satisfied. We conjecture that this condition is optimal in an information theoretic sense. Both the signal-to-noise ratio condition and the localization rates we obtain exhibit a quadratic dependence on the number of items to be compared, which matches the sample complexity bound for two sample testing for the \btl{} model recently derived by \cite{nihar_siva2020}. 

We emphasize that the change point setting we consider have not been previously studied and both our methodology and the corresponding theoretical guarantees appear to be the first contribution  of its kind in this line of work. 



\
\\
{\bf Related work}

Change point detection is a classical problem in statistics that dates back to 1940s \citep{wald1945, page1954}. Contributions in the 1980s established asymptotic theory for change point detection methods \citep{vostrikova1981,james1987,yao_au1989}. Most of the classical literature studied the univariate mean model. Recently with more advanced theoretical tools developed in modern statistics, more delicate analysis of change point detection came out in high-dimensional mean models \citep{jirak2015,aston2018,wang_samworth2018}, covariance models \citep{aue2009, avanesov2018, wang_yi_rinaldo_2021_cov}, high-dimensional regression models \citep{rinaldo2021cpd_reg_aistats, wang2021_jmlr}, network models \citep{wang2021network_cpd}, and temporally-correlated times series \citep{cho2015, preuss2015,chen2021cpdtimeseries,wang2022temporaldependence}.

Although change point detection has already been extensively studied in many different settings, little is known about the case of pairwise comparison data. \cite{Hohle2010} numerically study the CUSUM method for online change point detection in logit models and \btl{} models without giving theoretical guarantees.
We aim to fill the gap in the literature and propose a theoretically trackable approach that can optimally localize abrupt changes in the pairwise comparison data.


\section{Model and assumptions}
\label{sec:bt} 
Below we introduce the time-varying \btl{} model with piece-wise constant coefficients that we are going to study and the sampling scheme for collecting pairwise comparison data over time. 

Suppose there is a connected comparison graph $\mclG = \mclG([n],E)$ with edge set $E\subseteq E_{\rm full} := \{(i,j):1\leq i<j\leq n\}$. We assume throughout that data are collected as a time series indexed by $t \in [T]:=\{1,\ldots,T\}$ that, at each time point $t$, a single pairwise comparison among a collection of $n$ items is observed. The  distinct pair $(i_t, j_t) \in [n]^2$ of items to be compared at time $t$ is randomly chosen from the edge set $E$ of $\mclG$, independently over time. That is,
\begin{equation}
    \mathbb{P}(i_t = i, j_t = j) = \frac{1}{|E|}, \ \forall (i,j)\in E.
    \label{eq:random_graph}
\end{equation}
For each $t $, let $y_t\in \{0,1\}$  denote the outcome of the comparison between $i_t$ and $j_t$, where $y_t=1 $ indicates that $i_t $ beats $j_t$ in the comparison. We assume that $y_t$ follows the \btl{} model \eqref{neqn:bradley_terry_prob_succ}, i.e.
\begin{equation}
    \mathbb{P}_{\bbrtheta ^*(t)}[y_t = 1|(i_t,j_t) ] = \frac{e^{\theta^*_{i_t}(t)}}{e^{\theta^*_{i_t}(t)} + e^{\theta^*_{j_t}(t)}},
    \label{eq:original_bt}
\end{equation}
where $\bbrtheta^*(t) = (\theta^*_{1}(t),\ldots \theta^*_n(t))$ is, a possibly time-varying, parameter that belongs to the set
\begin{equation}
    \Theta_{B}:=\{\bbrtheta \in \mathbb{R}^n: \mathbf{1}_n^\top \bbrtheta=0,\ \|\bbrtheta\|_{\infty}\leq B\},
    \label{eq:Theta_B}
\end{equation}
for some $B>0$.
In the recent literature on the \btl{} model, the parameter $B$ is referred to as the {\it dynamic range} \citep[see, e.g.,][]{chen2019spectralregmletopk} which readily implies a bound on the smallest possible probability that an item is beaten by any other item. Indeed, it follows from \eqref{eq:original_bt} and \eqref{eq:Theta_B} that
\begin{equation}\label{eq:dynamic.range}
    \min_{t\in [T],i,j \in [n]}P_{ij}(t) \geq e^{-2B} / (1 + e^{-2B}) : = p_{lb} > 0.
\end{equation}

\bnrmk
The quantity $p_{lb}$ have appeared in several equivalent forms in the \btl{} literature, e.g., $\max_{i,j\in [n]}\frac{w^*_i}{w^*_j}$ \citep{simons1999,negahban2017rankcentralitypairwisecomparisons} and $e^{2B}$ \citep{li2022uai}.
The minimal winning probability $p_{lb}$ provides a way of quantifying the difficulty in estimating the model parameters, with a small $p_{lb}$ implying that some items are systematically better than others, a fact that is known to lead to non-existence of the MLE \citep[see, e.g.][]{ford1957} and to hinder parameter estimability. In the \btl{} literature  the dynamic range $B$ and, as a result, the quantity $p_{lb}$ are often treated as known constants and thus omitted \citep{shah2015estimationfrompairwisecomps, chen2020partialtopkranking}, a strong assumption  that results in  an implicit regularization but potentially hides an important feature of the model. As argued in \cite{bong2021generalized}, in high-dimensional settings this may not be realistic. We will allow for the possibility of a varying $B$ and $p_{lb}$, and keep track of the effect of  these parameters on our consistency rates. 
\enrmk


It is convenient to rewrite \eqref{eq:original_bt} in a different but equivalent form that is reminiscent of logistic regression and will facilitate our analysis. One can express the fact that, at time $t$, the items $i_t$ and $j_t$ are randomly selected from $\mclG([n],E)$ to be compared using a random $n$-dimensional vector $\bfx (t)$ that is drawn from the sets of all vectors in $\{-1, 0,1 \}^n$ with exactly two-non-zero entries of opposite sign, namely $x _{i_t}(t) = 1$ and $x _{j_t}(t) = -1$ for $(i_t,j_t)\in E$. Then equation \eqref{eq:original_bt} can be written as
\begin{equation}
\mathbb{P}_{\bbrtheta ^*(t)}[y_t = 1|\bfx (t) ] = \psi\left({ \bfx (t)^\top\bbrtheta^*(t)  }\right),
\label{eq:model_bt}
\end{equation} 
where $\psi(x) = \frac{1}{1 + e^{-x}}$ is the sigmoid function. For any time interval $\mclI \subset [T]$ we then assume that the data take the form of an i.i.d. sequence $\{(\bfx (t),y_t)\}_{t\in \mclI}$, where each $\bfx (t)$ is an i.i.d. draw from $\{-1, 0,1 \}^n$ with aforementioned properties and, conditionally on $\bfx (t)$, $y_t$ is a Bernoulli random variable with success probability \eqref{eq:original_bt}. The negative log-likelihood of the data is then given by
\begin{equation}
            L({\bbrtheta},\mclI) = \sum_{t\in \mclI}\ell_t(\bbrtheta),\ \text{where} \ \ell_t(\bbrtheta):=
    \ell(\bbrtheta; y_t, \bfx(t)) = -y_t\bfx (t)^\top {\bbrtheta} + \log [1 + \exp(\bfx (t)^\top {\bbrtheta})].
    \label{eq:likelihood}
\end{equation}
For a time interval $\mclI$, we can define a random \textit{comparison graph} $\mclG_{\mclI}(V_{\mclI}, E_{\mclI})$ with vertex set $V \defined [n]$ and edge set $E_{\mclI} \defined \{(i,j): i \text{ and } j \text{ are compared in }\mclI\}$. It is well-known that the topology of $\mclG_{\mclI}(V_{\mclI}, E_{\mclI})$ plays an important role in the estimation of \btl{} parameters \citep{shah2015estimationfrompairwisecomps}. Under assumption \eqref{eq:random_graph}, the comparison graph over $\mclI$ follows the random graph model $G([n], |\mclI|)$, which has $|\mclI|$ edges randomly picked from the edge set $E$ with replacement. Therefore, the process $\{(\bfx (t),y_t)\}_{t\in \mclI}$ is stationary as long as $\bbrtheta^*(t)$ is unchanged over $\mclI$.

In the change point \btl{} model we assume that, for some unknown integer $K \geq  1$, there exist $K+2$ points $\{\eta_k\}_{k=0}^{K+1}$ such that $1=\eta_0<\eta_1<\cdots<\eta_K<\eta_{K+1}=T$ and  $\bbrtheta^*(t)\neq \bbrtheta^*(t-1)$ whenever $t \in \{\eta_k\}_{k \in [K]}$. Define the \textit{minimal spacing} $\Delta$ between consecutive change points and the \textit{minimal jump size} $\kappa$ as
\begin{equation}
    \Delta = \min_{k\in [K+1]}(\eta_k -\eta_{k-1}), \quad \kappa = \min_{k\in [K+1]}\|\bbrtheta^*(\eta_k) - \bbrtheta^*(\eta_{k-1})\|_2.
\end{equation}
As we mentioned in the introduction, the goal of change point localization is to produce an estimator of the change points  $\{\hat{\eta}_k\}_{k\in [\hat{K}]}$ such that,  with high-probability as $T\rightarrow \infty$, we recover the correct number of change points and the localization error is a vanishing fraction of the minimal distance between change points, i.e. that
\begin{equation}
    \hat{K} = K, \text{ and } \max_{k\in [K]} |\hat{\eta}_k - \eta_k|/\Delta =o(1).
    \label{eq:consistency}
\end{equation}
In change point literature,  estimators satisfying the above conditions are called  {\it consistent.} In the next section we will present two change point estimators and prove their consistency.

\section{Main results}
\label{sec:main_result}
To estimate the  change points, we solve the following regularized maximum likelihood problem over all possible partitions $\mathcal{P}$ of the time course $[T]$:
\begin{equation}
  \hat{\mclP} = \argmin_{\mclP} \left\lbrace \sum_{\mclI\in \mclP} L(\hat{\bbrtheta}(\mclI),\mclI) + \gamma |\mclP| \right\rbrace, \quad \hat{\bbrtheta}(\mclI) = \argmin_{\bbrtheta \in \Theta_{B}} L(\bbrtheta,\mclI),
  \label{eq:P_hat_bt}
\end{equation}
where $L(\bbrtheta,\mclI)$ is the negative log-likelihood function for the \btl{} model defined in \eqref{eq:likelihood} and $\gamma > 0$ is an user-specified tuning parameter. Here a partition $\mclP$ is defined as a set of integer intervals:
\begin{equation}
  \mclP = \{[1,p_1),[p_1,p_2),\ldots,[p_{K_\mclP},T]\}, 1<p_1<p_2<\cdots<p_{K_{\mclP}}<T.
\end{equation}
With $\tilde{K} = K_{\hat{\mclP}} = |\hat{\mclP}|-1$, the estimated change points $\{\tilde{\eta}_k\}_{k\in \tilde{K}}$ are then induced by $\tilde{\eta}_k = \hat{p}_k$, $k\in [\tilde{K}]$. The optimization problem \eqref{eq:P_hat_bt} has an $\ell_0$-penalty, and can be solved by a dynamic programming algorithm described in \Cref{algo:DP} with $O(T^2\mclC(T))$ complexity \citep{friedrich2008complexitypenalizedmestimation, rinaldo2021cpd_reg_aistats}, where $\mclC(T)$ is the complexity of solving $\min_{\bbrtheta}L(\bbrtheta, [1,T])$. 

In this section, we will demonstrate that the estimator returned by  \Cref{algo:DP} is consistent. Towards that goal, we require the following signal-to-noise ratio condition involving the parameters $\Delta$, $\kappa$, $B$, $n$, the sample size $T$, and the topological property of the underlying comparison graph $\mclG([n], E)$.

\bnassum [Signal-to-noise ratio]
\label{assp:snr_bt}
Let $\{(\bfx (t),y_t)\}_{t\in [T]}$ be i.i.d. observations generated from model \eqref{eq:random_graph} and  \eqref{eq:model_bt}  with parameters $\{\bbrtheta^*(t)\}\subset \Theta_{B}$ defined in \eqref{eq:Theta_B}. We assume  that for  a diverging sequence $\{\mathcal{B}_T\}_{T \in \mathbb Z^+}$,
    \begin{equation}\label{eq:assp-bt}
      \Delta \cdot{\kappa^2} \geq \mathcal{B}_Tp_{lb}^{-4}{K}\frac{|E|nd_{\max}}{\lambda_2^2(L_{\mclG})}\log(Tn),
    \end{equation}
where we recall that $p_{ lb}:= \frac{e^{-2B}}{1 + e^{-2B}}$, $d_{\max}$ is the maximal degree of nodes in $\mclG$ and $\lambda_2(L_{\mclG})$ is the second smallest eigenvalue of the Laplacian of $\mclG$ \footnote{For a simple undirected graph $\mclG$ with (binary) adjacency matrix $A$, the Laplacian $L_{\mclG}:=D - A$ where $D= {\rm diag}(d_1,\cdots,d_n)$ where $d_i$ is the degree of node $i$.}.
\enassum
The formulation of signal-to-noise ratio conditions involving all the parameters of the model has become a staple of modern change point analysis literature. To provide some intuition, the term $\Delta \cdot\kappa^2$ is a proxy for the strength of the signal of change points in the sense that the localization and detection problems are expected to become easier, as the magnitude of the jumps and the spacing between change points increase. On the other hand, the right hand side of \Cref{eq:assp-bt} collects terms that impact negatively the difficulty of the problem: the smaller the minimal win probability $p_{lb}$ and the algebraic connectivity $\lambda_2(L_{\mclG})$, the larger the number of items $n$ to compare  and the number of change points $K$, the more difficult it is to estimate the change points.


\bnrmk[One the topology of $\mclG$]
When the comparison graph $\mclG$ is a \textit{complete graph}, we have $|E| = \frac{n(n-1)}{2}$, $d_{\max} = n-1$, $\lambda_2(L_{\mclG}) = n$, so the assumption becomes
\begin{equation}
      \Delta\cdot {\kappa^2} \geq \mathcal{B}_Tp_{lb}^{-4}{K}n^2\log(Tn).
      \label{assp:snr_bt.complete}
\end{equation}
In this case, the comparison graph $\mclG_{\mclI}([n],E_{\mclI})$ is random graph $G(n,m)$ that have $m$ edges sampled uniformly randomly with replacement. $G(n,m)$ is similar to an Erd\"os-R\'enyi graph that is commonly used in the ranking literature \citep{chen2019spectralregmletopk, chen2020partialtopkranking}. In this regard, our result, which directly reflects the impact of the general topology of the sampling graph, is fairly general and in line with recent advances in statistical ranking.

Also note that in general, $\lambda_2\leq \lambda_n\leq 2d_{\max}$, so the assumption \eqref{eq:assp-bt} ensures that the sample complexity $m\geq C_0 \frac{|E|\log n}{\lambda_2(L_\mclG)}$ in \Cref{lem:eig_laplacian_g} is satisfied in the worst case $\kappa^2\asymp n$.
\enrmk

\bnrmk[On the sharpness of the signal-to-noise ratio condition]
We will now argue that the requirement \eqref{assp:snr_bt} imposed by the signal-to-noise ratio (SNR for brevity) is reasonably sharp by relating it to the sample complexity of a two-sample testing problem. To that effect, consider the simplified setting in which there is only one change point at time $\Delta = T/2$ and $\mclG$ is a complete graph. In this case, it can be shown that the SNR condition \eqref{assp:snr_bt} becomes (see \Cref{prop:single_cp})
 \begin{equation}\label{eq:assp-bt.one.change.point}
      \Delta \cdot {\kappa^2} \geq \mathcal{B}_Tp_{lb}^{-2}{ n^2}\log(Tn),
    \end{equation}
i.e. the dependence on the dynamic range $B$ is through $p_{lb}^{-2}$ instead of $p_{lb}^{-4}$. 
It stands to reason that estimating the unknown change point $\Delta$ should be at least as hard as testing the null hypothesis that there exists a change point at time $\Delta$. Indeed, this testing problem should be easier because $\Delta$ has been revealed and because, in general, testing is easier than estimation. This can in turn be cast as a two-sample testing problem of the form
\begin{equation}\label{eq:two.sample}
  H_0: \bfP(\bbrtheta^{(1)}) = \bfP(\bbrtheta^{(2)}) \text{ v.s. } H_1: \frac{1}{n}\|\bfP(\bbrtheta^{(1)}) - \bfP(\bbrtheta^{(2)})\|_F\geq \epsilon,
\end{equation}
where $\epsilon > 0$ is to be specified, $\bbrtheta^{(1)}$ and $\bbrtheta^{(2)}$ are the \btl{} model parameters for the first and the last $\Delta$ observations respectively and, for $i\in \{1,2\}$, $\bfP(\bbrtheta^{(i)})$ is the $n \times n$ matrix of  winning probabilities corresponding to the \btl{} model parameter $\bbrtheta^{(i)}$ as specified by \eqref{eq:original_bt}. To see how one arrives at \eqref{eq:two.sample}, we have that, by \Cref{prop:prob_matrix_theta},
\begin{align}
\|\bfP(\bbrtheta^{(1)}) - \bfP(\bbrtheta^{(2)})\|^2_F \geq \frac{n p_{lb}^2}{16} \|\bbrtheta^{(1)} - \bbrtheta^{(2)}\|_2^2.
\end{align}
Thus, a change point setting with $ \|\bbrtheta^{(1)} - \bbrtheta^{(2)}\|_2^2 = \kappa^2$, translates into the testing problem \eqref{eq:two.sample} with $\epsilon^2 = \kappa^2 p^2_{lb}/(16 n )$. 
By Theorem 7 of \cite{nihar_siva2020}, there exists an algorithm that will return a consistent test for \eqref{eq:two.sample} based on two independent samples of size $N$ if $N\geq c{n^2\log(n)}\frac{1}{n\epsilon^2}$. When we apply this result to the simplified change point settings described above (by replacing  $N$ and $\epsilon^2$ with $\Delta$ and $\kappa^2 p^2_{lb}/(16 n)$ respectively) we conclude that  the sample complexity bound of Theorem 7 of \cite{nihar_siva2020} corresponds, up to constants, to the above SNR condition \eqref{eq:assp-bt.one.change.point} save for the terms $\log(T)$ and $\mathcal{B}_T$. Thus, we conclude that the assumed SNR condition for change point localization is essentially equivalent to the sample complexity needed to tackle the simpler two-sample testing problem, an indication that our assumption is sharp.

Finally, we take notice that, when there are multiple change points, in our analysis it appears necessary to strengthen the signal-to-noise ratio condition \eqref{eq:assp-bt.one.change.point} to \eqref{assp:snr_bt} by requiring a dependence on $p_{lb}^{-4}$.

\enrmk


We are now ready to present our first consistency result.

\bnthm
\label{thm:consistency_bt}
Let $\{\tilde{\eta}_k\}_{k\in [\tilde{K}]}$ be the estimates of change points from \Cref{algo:DP} with the tuning parameter $\gamma = C_{\gamma}p_{lb}^{-2}(K+1)\frac{nd_{\max}}{\lambda_2(L_{\mclG})}\log(Tn)$ where $C_{\gamma}$ is a universal constant. Under \Cref{assp:snr_bt} we have
\begin{equation}
  \mathbb{P}\left\lbrace \tilde{K} = K,\quad \max_{k\in [K]}|\tilde{\eta}_k - \eta_k|\leq C_P p_{lb}^{-4}{K}\frac{|E|nd_{\max}}{\kappa^2\lambda_2^2(L_{\mclG})}\log(Tn) \right\rbrace\geq 1 - 2(Tn)^{-2},
\end{equation}
where $C_P>0$ is a universal constant that depends on $C_{\gamma}$.
\enthm

\Cref{thm:consistency_bt} gives a high-probability upper bound for the localization error of the output  $\{\tilde{\eta}_k\}_{k\in [\tilde{K}]}$ of  \Cref{algo:DP}. By \Cref{assp:snr_bt}, it follows that as $T\rightarrow \infty$, with high probability,
\begin{equation}
    \max_{k\in [K]}|\tilde{\eta}_k - \eta_k|\leq C_P p_{lb}^{-4}{K}\frac{|E|nd_{\max}}{\kappa^2\lambda_2^2(L_{\mclG})}\log(Tn) \leq C_P \frac{\Delta}{\mathcal{B}_T} = o(\Delta),
    \label{eq:consistency_1}
\end{equation}
where we use the singal-to-noise ratio assumption $\Delta\cdot {\kappa^2} \geq \mathcal{B}_Tp_{lb}^{-4}{K}\frac{|E|nd_{\max}}{\lambda_2^2(L_{\mclG})}\log(Tn)$ in the last inequality and the fact that $\mathcal{B}_{T}$ diverges in the final step. This implies that  the estimators $\{\tilde{\eta}_k\}_{k\in [\tilde{K}]}$ are consistent. Moreover, when $K=0$ or there is no change point, it is guaranteed that, with high probability, \Cref{algo:DP} will return an empty set. We summarize this property as \Cref{prop:0_change} and include it in \Cref{sec:props} due to the limit of space.

\begin{algorithm}[!ht]
 \setstretch{1.15}
\textbf{INPUT:}  Data $\{ (\bfx(t), y_t) \}_{t\in [T]}$, tuning parameter $\gamma$.
\\
 	   Set $S= \emptyset$,  $\mathfrak{ p} =-\mathbf{1}_T $,  $\bfb =( \gamma, \infty , \ldots,  \infty)\in \mathbb{R}^T$.  Denote $b_i$ to be the $i$-th entry of $\bfb$. 
 	 \\
 	\For {$r  $ in $\{ 2,\ldots, T\} $}{
 	 \For {$l  $ in $\{ 1,\ldots,r - 1\}$}{
 	   $$ b \leftarrow b_{l}+ \gamma +   L (\hat{\bbrtheta}(\mclI),\mclI)  \quad \text{where} \quad \mclI=( l, \ldots, r]  ;$$
 	 \If {$b <  b_r $}{
 	    $b_r \leftarrow  b $; $\mathfrak{p}_r  \leftarrow l  $.
 	 }
 	}
 	}
  To compute  the change point estimates from $  \mathfrak{p}  \in \mathbb N^T $, 	
 $ k \leftarrow  T  $.
 \\
 	  \While {$k >1$}{
 	    $ h  \leftarrow \mathfrak{p}_ k  $  ; 
 	    $S = S \cup  h $; $ k  \leftarrow h $.}
 	
\textbf{OUTPUT:}  The estimated change points $S=\{\tilde{\eta}_k\}_{k\in \tilde{K}}$.  
 	
 \caption{Dynamic Programming. DP $(\{ ( \bfx (t), y_t) \}_{t\in [T]}, \gamma)$}
\label{algo:DP}
 
\end{algorithm}

\begin{algorithm}[!ht]
  \setstretch{1.15}

  \textbf{INPUT:} Data $\{(\bfx(t), y_t)\}_{t\in [T]}$, $\{\widetilde{\eta}_k\}_{k \in [\widetilde{K}]}$, $(\widetilde{\eta}_0, \widetilde{\eta}_{\widetilde{K} + 1}) \leftarrow (1, T)$.
  \\
  \For{$k = 1, \ldots, \widetilde{K}$}  {

\begin{equation}
    \begin{split}
    (s_k, e_k) &\leftarrow (2\widetilde{\eta}_{k-1}/3 + \widetilde{\eta}_{k}/3, \  \widetilde{\eta}_{k}/3 + 2\widetilde{\eta}_{k+1}/3);\\
    \hat{\eta}_k &\leftarrow \argmin_{\eta \in \{s_k + 1, \ldots, e_k - 1\}} \Bigg\{\min_{\bbrtheta^{(1)} \in \Theta_{B}}\sum_{t = s_k + 1}^{\eta}\ell_t(\bbrtheta^{(1)}) + \min_{\bbrtheta^{(2)} \in \Theta_{B}}\sum_{t = \eta + 1}^{e_k}\ell_t(\bbrtheta^{(2)})\Bigg\}; 
    \end{split}
    \label{eq-g-lasso}
\end{equation}
}
  \textbf{OUTPUT:} $\{\hat{\eta}_k\}_{k\in [\widetilde{K}]}$.
\caption{Local Refinement. }
\label{algo:local_refine_bt}
 
\end{algorithm} 

Inspired by previous works \citep{wang2021network_cpd,rinaldo2021cpd_reg_aistats}, we can further improve the localization error by applying a local refinement procedure as described in \Cref{algo:local_refine_bt}  to $\{\tilde{\eta}_k\}_{k\in [\tilde{K}]}$. This methodology takes as input any preliminary estimator of the change points that estimates the number of change points correctly with a localization error that is a (not necessarily vanishing) fraction of the minimal spacing $\Delta$, and returns a new estimator with a provably smaller localization error. A natural preliminary estimator is the one returned in \cref{algo:DP}. The next result derives the improved localization rates delivered by the local refinement step. The two improvements are the elimination of the term $K$ in the rate and a better dependence on $p_{lb}$.

\bnthm
\label{thm:rate_dp_local_refine}
Let $\{\hat{\eta}_k\}_{k\in [\hat{K}]}$ be the output of \Cref{algo:local_refine_bt}  with input $\{\tilde{\eta}_k\}_{k\in [\hat{K}]}$  returned by \Cref{algo:DP}. Under \Cref{assp:snr_bt}, for all sufficiently large $T$ we have
\begin{align}\label{eq:rate_dp_local_refine}
  \mathbb{P}\left\lbrace \hat{K} = K,\quad \max_{k\in [K]}|\hat{\eta}_k - \eta_k|\leq C_R p_{lb}^{-2}\frac{|E|nd_{\max}}{\kappa^2\lambda_2^2(L_{\mclG})}\log(Tn) \right\rbrace\geq 1 - 2(Tn)^{-2},
\end{align}
where $C_R>0$ is a universal constant that depends on $C_{\gamma}$.
\enthm
\bnrmk
By ``sufficiently large $T$'' in the theorem statement, we mean that $T$ should be large enough to make $\max_{k\in [K]}|\hat{\eta}_k - \eta_k|\leq \Delta/5$ (see \Cref{prop:local_refine} in \Cref{sec:appendix} for details). Such $T$ exists because of \Cref{eq:consistency_1} and the fact that $\mathcal{B}_T$ is diverging in $T$.
\enrmk
We conjecture that the rate \eqref{eq:rate_dp_local_refine} resulting from the local refinement procedure is, aside possibly from a logarithmic factor, minimax optimal.

\section{Experiments}
\label{sec:experiment}

In this section, we study the numerical performance of our newly proposed method based on a combination of dynamic programming with local refinement, which we will refer to as DPLR; see Algorithms \ref{algo:DP} and \ref{algo:local_refine_bt}. We note that the detection of multiple change points in pairwise comparison data has not been studied before, as \cite{Hohle2010} only focus on single change point detection for pairwise comparison data, so we are not aware of any existing competing methods in the literature. Thus, we develop a potential competitor based on the combination of \textit{Wild Binary Segmentation} (WBS) \citep{Fryzlewicz2014}, a popular method for univariate change point detection, and the likelihood ratio approach studied in \cite{Hohle2010}. We will call this potential competitor WBS-GLR (GLR stands for generalized likelihood ratio). Due to the limit of space, we include the detail of WBS-GLR in \Cref{sec:wbs-glr}, and results of additional experiments in \Cref{sec:additional_experiment}, where additional settings are considered. Furthermore, we discuss and compare the performance of two other potential competitors in \Cref{sec: other competitor}. 

All of our simulation results show that our proposed method DPLR outperforms WBS-GLR in the sense that DPLR gives more accurate change point estimates with similar running time. Each experiment is run on a virtual machine of Google Colab with Intel(R) Xeon(R) CPU of 2 cores 2.30 GHz and 12GB RAM. 
All of our reproducible code is openly accessible
\footnote{Code repository: \url{https://github.com/MountLee/CPD_BT}}

\paragraph{Simulation Settings.} Suppose we have $K$ change points $\{ \eta_k\}_{k\in [K]} $ in the sequential pairwise comparison data, with $\eta_0 =1 $. We can use $\bbrtheta^*(\eta_k)$ to represent the value of true parameters after the change point $\eta_k$. 
To begin, we define $\theta_i^*(\eta_0) $ as follows. 
For $1< i \le n$, we set $\theta_i^*(\eta_0) = \theta_1^*(\eta_0) +(i -1)\delta$ with some constant $\delta$. In each experiment, we set $\delta$ first and then set $\theta_1^*(\eta_0)$ to make $\mathbf{1}^\top_n\bbrtheta^*(\eta_0) = 0$. For a given $n$, we set $\delta = \frac{1}{n - 1}\psi^{-1}(p) =\frac{1}{n - 1}\log(\frac{p}{1 - p})$ where $\psi^{-1}$ is the inverse function of $\psi$ and $p = 0.9$. Recall that $P_{ij} = \psi(\theta_i-\theta_j)$ is the winning probability, so the value of $\delta$ guarantees that the maximum winning probability is 0.9. We consider three types of changes:

Type I (reverse): $\theta_i^*(\eta_k) = \theta^*_{n + 1 - i}(\eta_0)$.

Type II (block-reverse): $\theta_i^*(\eta_k) = \theta^*_{[\frac{n}{2}] + 1 - i}(\eta_0)$ for $i\leq [\frac{n}{2}]$; $\theta_i^*(\eta_k) = \theta^*_{[\frac{n}{2}] + n + 1 - i}(\eta_0)$ for $i> [\frac{n}{2}]$.

Type III (block exchange): $\theta_i^*(\eta_k) = \theta^*_{i + [\frac{n}{2}]}(\eta_0)$ for $i\leq [\frac{n}{2}]$; $\theta_i^*(\eta_k) = \theta^*_{i - [\frac{n}{2}]}(\eta_0)$ for $i> [\frac{n}{2}]$.

We consider four simulation settings. For each setting, we set the comparison graph $\mclG([n],E)$ to be the complete graph and $T = (K+1)\Delta$ with true change points located at $\eta_i = i \Delta$ for $i\in [K]$. To describe the true parameter at each change point, we use an ordered tuple. For instance, (I, II, III, I) means that $K=4$ and the true parameters at $\eta_1,\eta_2,\eta_3, \eta_4$ are determined based on $\bbrtheta^*(\eta_0)$ and the change type I, II, III, and I, respectively.

\begin{table}[!h]
 \centering
 \begin{tabular}{m{2cm}m{2cm}m{2cm}C{1.5cm}C{1.5cm}C{1.5cm}}
  \hline
  & $H(\hat{\eta},\eta)$ & Time & $\hat{K}<K$ &$\hat{K}=K$ &$\hat{K}>K$
 \\
 \hline
 \multicolumn{6}{c}{{\bf Setting (i)} \ \ $n = 10, K = 3, \Delta = 500$, Change (I, II, III)} \\
 DPLR & 9.2 (9.1) & 49.7s (0.7) & 0 & 100 & 0 \\
 WBS-GLR & 15.2 (7.9) & 31.9s (3.9) & 0 & 100 & 0 \\
 \hline
 \multicolumn{6}{c}{ {\bf Setting (ii)} \ \ $n = 20, K = 3, \Delta = 800$, Change (I, II, III)} \\
 DPLR & 9.0 (9.9) & 118.5s (2.2) & 0 & 100 & 0 \\
 WBS-GLR & 240.5 (220.3) & 144.2s (12.5) & 0 & 40 & 60 \\
 \hline
 \multicolumn{6}{c}{{\bf Setting (iii)}\ \ $n = 100, K = 2, \Delta = 1000$, Change (I, II)} \\
 DPLR & 13.4 (14.4) & 167.4s (3.3) & 0 & 100 & 0 \\
 WBS-GLR & 111.9 (195.6) & 215.9s (17.0) & 0 & 79 & 21 \\
 \hline
 \multicolumn{6}{c}{{\bf Setting (iv)} \ \ $n = 100, K = 3, \Delta = 2000$, Change (I, II, III)} \\
 DPLR & 12.4 (12.1) & 402.4s (7.4) & 0 & 100 & 0 \\
 WBS-GLR & 412.3 (495.5) & 400.0s (40.9) & 0 & 57 & 43 \\
 \hline
 \end{tabular}
 \caption{Comparison of DPLR and WBS-GLR under four different simulation settings. 100 trials are conducted in each setting. For the localization error and running time (in seconds), the average over 100 trials is shown with standard error in the bracket. The three columns on the right record the number of trials in which $\hat{K}<K$, $\hat{K}=K$, and $\hat{K}>K$ respectively.}
 \label{tab:compare_methods}
\end{table}

For the constrained MLE in \Cref{eq:P_hat_bt}, we use the function in \texttt{sklearn} for fitting the $\ell_2$-penalized logistic regression, as it is well-known that the constrained and the penalized estimators for generalized linear models are equivalent. For both DPLR and WBS-GLR, we use $\lambda = 0.1$. For $M$, the number of random intervals in WBS-GLR, we set it to be 50 as a balance of time and accuracy. 

For both methods, we use cross-validation to choose the tuning parameter $\gamma$. Given the sequential pairwise comparison data in each trial, we use samples with odd time indices as training data and even time indices as test data. For each tuning parameter, the method is applied to the training data to get estimates of change points. Then a \btl{} model is fitted to the test data for each interval determined by the estimated change points. The tuning parameter and the corresponding change point estimators with the minimal test error (negative loglikelihood) are selected. We run 100 trials for each setting.

\paragraph{Results.} To measure the localization errors, we use the Hausdorff distance $H(\{\hat{\eta}_i\}_{i\in [\hat{K}]},\{\eta_i\}_{i\in [K]})$ between the estimated change points $\{\hat{\eta}_i\}_{i\in [\hat{K}]}$ and the true change points $\{\eta_i\}_{i\in [K]}$. The Hausdorff distance $H(S_1, S_2)$ between two sets of scalars is defined as
\begin{align} H(S_1, S_2) = \max\{\sup_{x\in S_1}\inf_{y\in S_2}|x - y|, \sup_{y\in S_2}\inf_{x\in S_1}|x - y|\}.
\end{align}
The results are summarized in \Cref{tab:compare_methods}, where we use $H(\hat{\eta},\eta)$ to denote the localization error for brevity. As we can see, our proposed method DPLR gives more accurate localization with similar running time compared to the potential competitor WBS-GLR.

\section{Application: the National Basketball Association games}
\label{sec:application}

\begin{table}[ht!]
\centering
\scalebox{0.9}{
\begin{tabular}{lrlrlrlr}
\hline
\hline
\multicolumn{2}{c}{\textbf{S1980-S1985}}    & \multicolumn{2}{c}{\textbf{S1986-S1991m}}   & \multicolumn{2}{c}{\textbf{S1991m-S1997}}   & \multicolumn{2}{c}{\textbf{S1998-S2003}}    \\
\hline
Celtics              & 1.1484               & Lakers               & 1.1033               & Bulls                & 0.9666               & Spurs                & 0.8910               \\
76ers                & 0.9851               & Pistons              & 0.7696               & Jazz                 & 0.8618               & Lakers               & 0.8744               \\
Bucks                & 0.7828               & Celtics              & 0.7304               & Knicks               & 0.5908               & Kings                & 0.6833               \\
Lakers               & 0.7779               & Trail Blazers        & 0.6848               & Suns                 & 0.5628               & Mavericks            & 0.5087               \\
Nuggets              & 0.0789               & Bulls                & 0.6647               & Rockets              & 0.5032               & Trail Blazers        & 0.4899               \\
Trail Blazers        & 0.0636               & Jazz                 & 0.5179               & Spurs                & 0.4742               & Jazz                 & 0.3944               \\
Suns                 & 0.0636               & Bucks                & 0.3474               & Trail Blazers        & 0.4176               & Timberwolves         & 0.3913               \\
Spurs                & 0.0611               & Suns                 & 0.3472               & Cavaliers            & 0.3751               & Pacers               & 0.3165               \\
Nets                 & 0.0215               & Rockets              & 0.3156               & Magic                & 0.3009               & Hornets              & 0.1002               \\
Pistons              & -0.0252                   & 76ers                & 0.2195               & Lakers               & 0.2730                & 76ers                & 0.0993               \\
Knicks               & -0.1333                   & Cavaliers            & 0.1885               & Pacers               & 0.2688               & Suns                 & 0.0721               \\
Rockets              & -0.1950                   & Mavericks            & 0.1798               & Hornets              & 0.2465               & Pistons              & 0.0249               \\
Jazz                 & -0.2926                   & Knicks               & 0.0583               & Heat                 & 0.1445               & Bucks                & -0.0146              \\
Kings                & -0.3104              & Warriors             & 0.0441               & Pistons              & -0.2028              & Rockets              & -0.0525              \\
Mavericks            & -0.3104              & Spurs                & 0.0035               & Nets                 & -0.2122              & Knicks               & -0.1420               \\
Bulls                & -0.3115               & Nuggets              & -0.0232              & Warriors             & -0.3075              & Heat                 & -0.1455              \\
Warriors             & -0.4330              & Pacers               & -0.0237              & Celtics              & -0.3288              & Nets                 & -0.2276              \\
Pacers               & -0.5500              & Kings                & -0.7006              & Kings                & -0.4808              & Magic                & -0.2650               \\
Clippers             & -0.6443              & Nets                 & -0.7666              & Clippers             & -0.5419              & Celtics              & -0.2885              \\
Cavaliers            & -0.7771              & Clippers             & -0.7788              & Bucks                & -0.5864              & Nuggets              & -0.4894              \\
Heat                 &NA               & Magic                & -0.8969              & Nuggets              & -0.6272              & Clippers             & -0.6250               \\
Hornets              &NA                & Timberwolves         & -0.9554              & Timberwolves         & -0.6570               & Cavaliers            & -0.6796              \\
Magic                &NA              & Heat                 & -0.9874              & 76ers                & -0.8869              & Warriors             & -0.7362              \\
Timberwolves         &NA              & Hornets              & -1.0418              & Mavericks            & -1.1542              & Bulls                & -1.1801              \\
\hline
\hline
\multicolumn{2}{c}{\textbf{S2004-S2006}}    & \multicolumn{2}{c}{\textbf{S2007-S2009}}    & \multicolumn{2}{c}{\textbf{S2010-S2012}}    & \multicolumn{2}{c}{\textbf{S2013-S2015}}    \\
\hline
Spurs                & 1.0532               & Lakers               & 1.0097               & Heat                 & 0.9909               & Warriors             & 1.3617               \\
Suns                 & 0.9559               & Celtics              & 0.8699               & Spurs                & 0.8653               & Spurs                & 1.2728               \\
Mavericks            & 0.9338               & Magic                & 0.7741               & Bulls                & 0.8292               & Clippers             & 0.9909               \\
Pistons              & 0.8120                & Cavaliers            & 0.7466               & Nuggets              & 0.5857               & Rockets              & 0.6158               \\
Heat                 & 0.2713               & Spurs                & 0.6270                & Lakers               & 0.4922               & Trail Blazers        & 0.5501               \\
Rockets              & 0.1803               & Mavericks            & 0.5686               & Mavericks            & 0.4121               & Mavericks            & 0.4197               \\
Cavaliers            & 0.1510                & Jazz                 & 0.5169               & Clippers             & 0.3413               & Cavaliers            & 0.3872               \\
Nuggets              & 0.1322               & Nuggets              & 0.4751               & Celtics              & 0.2901               & Heat                 & 0.3215               \\
Kings                & 0.0542               & Suns                 & 0.4146               & Knicks               & 0.1990                & Pacers               & 0.3202               \\
Lakers               & 0.0166               & Hornets              & 0.3593               & Pacers               & 0.1233               & Bulls                & 0.2104               \\
Nets                 & -0.0149              & Rockets              & 0.3428               & Rockets              & 0.1227               & Hornets              & 0.0145               \\
Timberwolves         & -0.0566              & Trail Blazers        & 0.2750                & Jazz                 & 0.0167               & Pistons              & -0.1710               \\
Clippers             & -0.0646              & Bulls                & -0.1260               & Trail Blazers        & -0.0549              & Suns                 & -0.1787              \\
Bulls                & -0.0680               & Pistons              & -0.1821              & Magic                & -0.0899              & Jazz                 & -0.1936              \\
Pacers               & -0.0824              & Heat                 & -0.2939              & Warriors             & -0.1402              & Celtics              & -0.2037              \\
Jazz                 & -0.1039              & 76ers                & -0.3418              & 76ers                & -0.1930               & Nets                 & -0.3093              \\
Magic                & -0.2482              & Warriors             & -0.3729              & Bucks                & -0.2362              & Nuggets              & -0.3140               \\
Warriors             & -0.2803              & Pacers               & -0.3936              & Suns                 & -0.3228              & Kings                & -0.4066              \\
76ers                & -0.3030               & Bucks                & -0.5456              & Nets                 & -0.4589              & Bucks                & -0.4516              \\
Celtics              & -0.5144              & Kings                & -0.7977              & Hornets              & -0.4670               & Timberwolves         & -0.6266              \\
Hornets              & -0.5641              & Knicks               & -0.8568              & Timberwolves         & -0.6034              & Magic                & -0.6398              \\
Bucks                & -0.6555              & Nets                 & -0.8935              & Kings                & -0.6929              & Knicks               & -0.6591              \\
Knicks               & -0.7101              & Clippers             & -1.0853              & Pistons              & -0.7807              & Lakers               & -0.9431              \\
Trail Blazers        & -0.8947              & Timberwolves         & -1.0901              & Cavaliers            & -1.2285              & 76ers                & -1.3676         \\
\hline
\end{tabular}
}
\caption{Fitted $\hat{\bbrtheta}$ (rounded to the fourth decimal) for 24 selected teams in seasons 1980-2016 of the National Basketball Association. Teams are ranked by the MLE $\hat{\bbrtheta}$ on subsets splitted at the estimated change points given by our DPLR method. S1980 means season 1980-1981 and S1991m means the middle of season 1991-1992. Heat(1988), Hornets(1988), Magic(1989), and Timberwolves(1989) were founded after S1985, so the corresponding entries are marked as NA.}
\label{tab:nba_2}
\end{table}

We study the game records of the National Basketball Association (NBA) \footnote{\url{https://gist.github.com/masterofpun/2508ab845d53add72d2baf6a0163d968}}. Usually a regular NBA season begins in October and ends in April of the next year, so in what follows, a season is named by the two years it spans over. The original data contains all game records of NBA from season 1946-1947 to season 2015-2016. We focus on a subset of 24 teams founded before 1990 and seasons from season 1980-1981 to season 2015-2016. 
All code of analysis is available online with the data
\footnote{Code repository: \url{https://github.com/MountLee/CPD_BT}}

We start with an exploratory data analysis and the results show strong evidence for multiple change points \footnote{Due to the limit of space, we include these results in \Cref{sec:additional_application}.}. Therefore, we apply our method DPLR to the dataset to locate those change points. We use the samples with odd time indices as training data and even time indices as test data, and use cross-validation to choose the tuning parameter $\gamma$.

To interpret the estimated change points, we fit the \btl{} model on each subset splitted at change point estimates separately. The result is summarized in \Cref{tab:nba_2}. Several teams show significant jumps in the preference scores and rankings around change points. Apart from this quantitative assessment, the result is also firmly supported by memorable facts in NBA history, and we will name a few here. In 1980s, Celtics was in the ``Larry Bird'' era with its main and only competitor ``Showtime'' Lakers. Then starting from 1991, Michael Jordan and Bulls created one of the most famous dynasties in NBA history. 1998 is the year Michael Jordan retired, after which Lakers and Spurs were dominating during 1998-2009 with their famous cores ``Shaq and Kobe'' and ``Twin Towers''. The two teams together won 8 champions during these seasons. S2010-S2012 is the well-known ``Big 3'' era of Heat. Meanwhile, Spurs kept its strong competitiveness under the lead of Timothy Duncan. From 2013, with the arise of super stars Stephen Curry and Klay Thompson, Warriors started to take the lead.

\section{Conclusions}
\label{sec:conclusion}

We have formulated and investigate a novel change point analysis problem for pairwise comparison data based on a high-dimensional \btl{} model. We have developed a novel methodology that yields consistent estimators of the change points, and establish theoretical guarantees with nonasymptotic localization error. To the best of our knowledge, this is the first work in the literature that addresses in both a methodological and theoretically sound way multiple change points in ranking data.  

Although we filled a big gap in the literature, there remain many open and interesting problems for future work. First,  we only consider pairwise comparison data modeled by the \btl{} model. Of course, there are other popular ranking models for general ranking data, e.g., the Plackett-Luce model\citep{Luce1959,Plackett1975}, Stochastically Transitive models\citep{shah2017stmodelspairwisecomps}, and the Mallows model \citep{tang19mallows}. It would be interesting to see that for those models how different the method and theory would be from our settings. We present some exploratory results on this in \Cref{sec: other competitor}. Second, we have focused on \textit{retrospective} setting of change point detection and \textit{passive} setting of ranking. On the other hand, \textit{online} change point detection \citep{Vovk2021Testing} and \textit{active ranking} \citep{heckel2019activerankingpairwisecomps, ren2021activeranking} are widely used in practice. Thus, it would be interesting to consider the online or active framework in change point detection for ranking data. Third, in the recent change point detection literature, incorporating temporal dependence is of growing interest \citep{chen2021cpdtimeseries,wang2022temporaldependence}, so investigating how temporal dependence in the pairwise comparison data can affect our results seems like a worthwhile direction. 


At last, we discuss potential societal impacts of our work. The \btl{} model does have
applications with potentially undesirable societal impacts, e.g., sports-betting \citep{mchale2011btforecasttennis},
which could amplify the negative impacts
of gambling. We recommend using our method for    research purposes  rather than gambling-driven purposes.

\noindent{\bf Acknowledgments}\label{subsec:acknowledgments}

We would like to thank the anonymous
reviewers for their feedback which greatly helped improve our exposition. Wanshan Li and Alessandro Rinaldo acknowledge partial support from NSF grant DMS-EPSRC 2015489.


\clearpage
\bibliographystyle{apalike}
\bibliography{refs}





\clearpage
\appendix

\begin{center}
\textbf{\LARGE Appendix of ``Detecting Abrupt Changes in Sequential Pairwise Comparison Data''}
\end{center}

This is the appendix of the paper ``Detecting Abrupt Changes in Sequential Pairwise Comparison Data'' as a supplementary material. It contains two parts:
\begin{enumerate}
    \item \Cref{sec:appendix_numerical} for some supplements to numerical results in Sections \ref{sec:experiment} and \ref{sec:application}.
    \item \Cref{sec:appendix} for the proof of main results and some additional propositions used in the main text.
\end{enumerate}

\section{Appendix: supplementary to numerical results}
\label{sec:appendix_numerical}
\subsection{Wild binary segmentation based on likelihood}
\label{sec:wbs-glr}
\textit{Binary segmentation} is a classical and popular method for detecting change points that can at least date back to \cite{Scott1974}. It is based on the so-called CUSUM statistics. In the case where we are interested in detecting the change point in the mean of univariate random variables $\{Y_t\}_{t\in [T]}$, the CUSUM statistic at time $t$ over an interval $(s,e)$ is defined as
\begin{equation}
    {\rm CUSUM}(t;s,e) := |\sqrt{\frac{e - t}{(e - s)(t - s)}}\sum_{i = s+1}^t Y_i - \sqrt{\frac{t - s}{(e - s)(e - t)}}\sum_{i = t+1}^e Y_i|.
\end{equation}
It is known that Binary Segmentation is consistent but not optimal (\cite{Venkatraman1992}). As an improvement, \cite{Fryzlewicz2014} propose Wild Binary Segmentation and show that it has a better localization rate.

\begin{algorithm}[H]
 \setstretch{1.15}
 	\textbf{INPUT:} Independent samples $\{Z_i\}_{i\in [n]}$, collection of intervals $\{ (\alpha_m,\beta_m)\}_{m\in [M]}$, tuning parameters $\gamma > 0$.
 
	\For{$m = 1, \ldots, M$}  {
		  $(s_m, e_m) \leftarrow [s,e]\cap [\alpha_m,\beta_m]$
		  \\
		\uIf{$e_m - s_m > 1$}{
			  $b_{m} \leftarrow \argmax_{s_m + 1 \leq t \leq e_m - 1}  \mclR(t;s_m,e_m)$
			 \\ $a_m \leftarrow \mclR(b_m;s_m,e_m)$}
		\Else {
		  $a_m \leftarrow -1$	} 
		
} 
$m^* \leftarrow \argmax_{m \in [M]} a_{m}$
\\
\If{$a_{m^*} > \gamma$}{
		  add $b_{m^*}$ to the set of estimated change points\\
		  WBS$((s, b_{m*}),\{ (\alpha_m,\beta_m)\}_{m\in [M]}, \gamma)$\\
		  WBS$((b_{m*}+1,e),\{ (\alpha_m,\beta_m)\}_{m\in [M]},\gamma ) $
}
		\textbf{OUTPUT:} The set of estimated change points.
\caption{Wild Binary Segmentation. WBS$((s, e),$ $\{ (\alpha_m,\beta_m)\}_{m\in [M]}, \gamma $)}
\label{algorithm:WBS}
 \end{algorithm} 
 
\Cref{algorithm:WBS} shows the general framework of WBS algorithm. For univariate mean, we have $\mclR(t;s,e) = {\rm CUSUM}(t;s,e)$. While for our problem, the Bradley-Terry model, we set $\mclR(t;s,e)$ to be the (logarithmic) generalized likelihood ratio given by
\begin{equation}
    \mclR(t;s,e) = GLR(t;s,e) := \max_{\bbrtheta_l\in \Theta_B}  \{-L(\bbrtheta_l,[s, t))\} + \max_{\bbrtheta_r\in \Theta_B} \{-L(\bbrtheta_r,[t,e])\} - L_{s,e},
    \label{eq:R_wbs_bt}
 \end{equation}
where $L_{s,e}:=\max_{\bbrtheta\in \Theta_B}\{-L(\bbrtheta,[s,e])\}$ and $L(\bbrtheta,\mclI)$ is the negative log-likelihood function over interval $\mclI$, as is defined in \Cref{eq:likelihood}. The use of generalized likelihood ratio in change point detection has been demonstrated in many previous works \citep{Hohle2010, wang2018univariate}. In fact, when $\{Y_t\}_{t\in [T]}$ follows Gaussian distribution with known variance, the GLR statistic at $t$ is the square of ${\rm CUSUM}(t;s,e)$.

Similar to the DP approach, WBS also has a tuning parameter $\gamma$. By \Cref{eq:R_wbs_bt} and the design of \Cref{algo:DP} and \ref{algorithm:WBS}, we know that the $\gamma$ parameters for both DP and WBS-GLR act as the threshold for the GLR statistic. Therefore, one should use the same candidate list of $\gamma$ for both methods when tuning parameters by cross-validation for fair comparison, as we do in all experiments.

In addition, the number of intervals $M$ acts as another tuning parameter and makes WBS more tricky to apply compared to the DP approach. In practice, people usually set intervals $\{ (\alpha_m,\beta_m)\}_{m\in [M]}$ to be uniformly randomly sampled from $[0,T]$. Although it doesn't affect the theoretical guarantee too much \cite{wang2018univariate}, numerically the performance of WBS heavily depends on $M$. Typically, the larger $M$ is, the more accurate the result is, and the more time it takes to execute WBS. When the model of the data is simple, e.g., univariate mean model, computation of $\mclR(t;s,e)$ is cheap and one can just set $M$ to be large to improve the localization accuracy. However, for more complex models like the \btl{} model, a large $M$ may not be computationally affordable, so it can be hard to set an appropriate value for $M$.

\clearpage
\subsection{Additional simulated experiments}
\label{sec:additional_experiment}
In \Cref{sec:experiment},  we consider simulation settings where both the signals $\bbrtheta^*(t)$ and changes of $\bbrtheta^*(t)$ at change points are set in a deterministic way.  In this section, we consider experiments where entries of $\bbrtheta^*(t)$ are randomly sampled  and   are randomly permuted at each change point. Suppose we have $K$ change points $\{ \eta_k\}_{k\in [K]} $ in the sequential pairwise comparison data, with $\eta_0 =1 $. We  use $\bbrtheta^*(\eta_k)$ to represent the value of true parameters after the change point $\eta_k$. 

To begin, we   set $ \{ \theta_i^*(\eta_0) \}_{i=1}^ n \overset{i.i.d.}{\sim} {\rm Uniform}[0,1]$. 
We further rescale $\bbrtheta^*(t)$ by setting 
$\theta_i^*(\eta_0) \leftarrow \frac{\psi^{-1}(0.9)}{\max_i \theta_i^*(\eta_0) - \min_i \theta_i^*(\eta_0)}\theta_i^*(\eta_0)$   
and then set  $\theta_i^*(\eta_0) \leftarrow \theta_i^*(\eta_0) - {\rm avg}(\bbrtheta^*(\eta_0))$. Here $\psi^{-1}(p) = \log(\frac{p}{1 - p})$ is the inverse function of $\psi$. Recall that $P_{ij} = \psi(\theta_i-\theta_j)$ is the winning probability. So by rescaling  $\bbrtheta^*(t)$,  we  guarantee that at time $\eta_0$, the maximum winning probability is 0.9. 

For each change point $\eta_k$, $k\geq 1$, we  randomly sample a permutation $\pi:[n]\mapsto [n]$ from the collection of all $n$-permutations and set $\theta_i^*(\eta_k) = \theta_{\pi(i)}^*(\eta_{k-1})$ for $i\in [n]$. We consider the same settings for $(n, K, \Delta)$ with the same tuning parameters as in \Cref{sec:experiment}, and summarize our new simulation results in \Cref{tab:compare_methods_random_change}

\begin{table}[!h]
 \centering
 \begin{tabular}{m{2cm}m{2cm}m{2cm}C{1.5cm}C{1.5cm}C{1.5cm}}
  \hline
  & $H(\hat{\eta},\eta)$ & Time & $\hat{K}<K$ &$\hat{K}=K$ &$\hat{K}>K$
 \\
 \hline
 \multicolumn{6}{c}{{\bf Setting (i)} \ \ $n = 10, K = 3, \Delta = 500$, Random change} \\
 DPLR & 12.1 (13.3) & 62.4s (2.1) & 0 & 100 & 0 \\
 WBS-GLR & 94.9 (174.8) & 33.6s (5.4) & 0 & 100 & 0 \\
 \hline
 \multicolumn{6}{c}{ {\bf Setting (ii)} \ \ $n = 20, K = 3, \Delta = 800$, Random change} \\
 DPLR & 23.9 (27.6) & 105.8s (4.2) & 0 & 100 & 0 \\
 WBS-GLR & 251.7 (219.9) & 133.7s (14.7) & 0 & 40 & 60 \\
 \hline
 \multicolumn{6}{c}{{\bf Setting (iii)}\ \ $n = 100, K = 2, \Delta = 1000$, Random change} \\
 DPLR & 43.1 (103.4) & 196.9s (3.9) & 1 & 99 & 0 \\
 WBS-GLR & 133.0 (194.9) & 210.0s (16.6) & 0 & 76 & 24 \\
 \hline
 \multicolumn{6}{c}{{\bf Setting (iv)} \ \ $n = 100, K = 3, \Delta = 2000$, Random change} \\
 DPLR & 28.3 (26.5) & 453.6s (9.2) & 0 & 100 & 0 \\
 WBS-GLR & 459.4 (512.8) & 410.5s (48.7) & 0 & 53 & 47 \\
 \hline
 \end{tabular}
 \caption{Comparison between DPLR and WBS-GLR under four different simulation settings with random   signals. For the localization error    and running time (in seconds), the averages over 100 trials are reported with standard errors in the brackets. The last three columns on the right record the number of trials in which $\hat{K}<K$, $\hat{K}=K$, and $\hat{K}>K$ respectively.}
 \label{tab:compare_methods_random_change}
\end{table}



In what follows, we further investigate the effect of signal strength by restricting the random permutation at each change point to a subset of $[n]$, and analyze the performance of both methods while varying the size of permuted subsets. The results are summarized in \Cref{tab:signal_strength}, where 50\% random permutation means at each change point $\eta_k$, only   50\% of  the entries of $\bbrtheta^*(\eta_{k-1})$ are randomly selected and  permuted to form $\bbrtheta^*(\eta_{k})$. Note that  as the proportion of the randomly  permuted  entries  increases, the random   perturbation strength   raises at the  change points.   As shown in  \Cref{tab:signal_strength}, our algorithm DPLR is able to provide more accurate   change point estimations as  the  random   perturbation strength increases.  

\begin{table}[H]
 \centering
 \begin{tabular}{m{1.8cm}m{1.2cm}m{2.2cm}m{2cm}C{1.3cm}C{1.3cm}C{1.3cm}}
 \hline
 \multicolumn{7}{c}{$n = 20, K = 3, \Delta = 800$} \\
 Random\ \ \ permutation & Method & $H(\hat{\eta},\eta)$ & Time & $\hat{K}<K$ &$\hat{K}=K$ &$\hat{K}>K$
 \\
 \hline
 50\% & DPLR & 362.8 (502.2) & 97.1s (10.4) & 27 & 67 & 6  \\
    & WBS & 407.5 (336.8) & 137.2s (21.7) & 10 & 21 & 69  \\
\hline
 75\% & DPLR & 114.4 (251.3) & 120.4s (4.4) & 8 & 91 & 1    \\
    & WBS & 349.6 (261.8) & 141.8s (17.2) & 13 & 28 & 59    \\
\hline
 100\% & DPLR & 23.9 (27.6) & 105.8s (4.2) & 0 & 100 & 0 \\
    & WBS & 251.7 (219.9) & 133.7s (14.7) & 0 & 40 & 60 \\
 \hline
 \end{tabular}
 \caption{Performance of DPLR and WBS-GLR under different signal strength. For the localization error and running time (in seconds), the average over 100 trials is shown with standard error in the bracket.}
 \label{tab:signal_strength}
\end{table}

\clearpage
\subsection{Additional results for real data applications}
\label{sec:additional_application}

\subsubsection{Exploratory analysis}

\begin{figure}[H]
    \centering
    \includegraphics[width = 0.95 \textwidth]{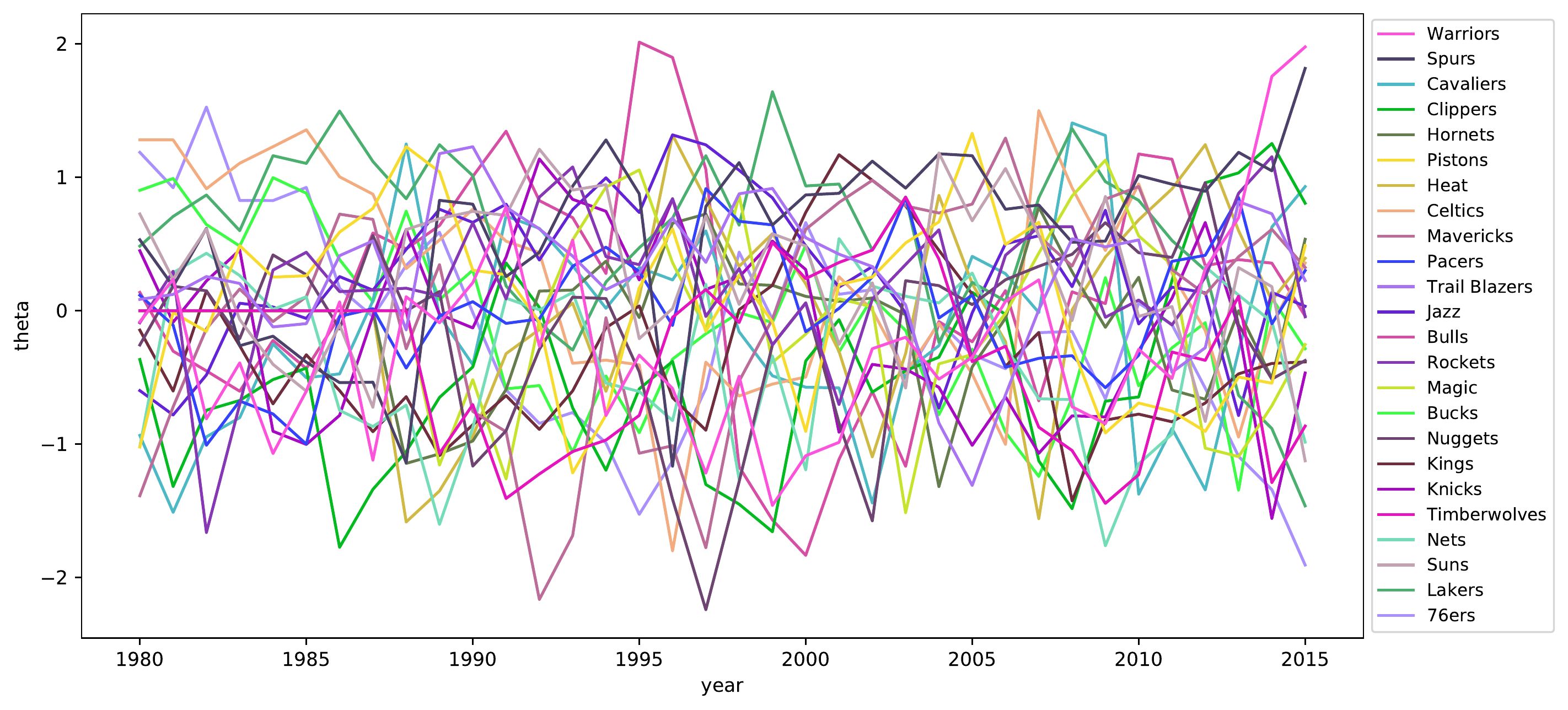}
    \caption{Path of $\hat{\bbrtheta}(\mclI_s)$ for $\mclI_s$ being each season of the NBA data.}
    \label{fig:nba_theta_path_2}
\end{figure}

We start our analysis by fitting the \btl{} model on each season and drawing the path of fitted $\hat{\bbrtheta}(\mclI_s)$, where $\mclI_s$ is the index interval for games in the $s$-th season in our range of interest, i.e., from season 1980-1981 to season 2015-2016. The resulting paths shown in \Cref{fig:nba_theta_path_2} are fairly noisy for interpretation and inference, and this is  a strong evidence that the data is unstationary. In addition, these unstructured paths explain why we need some principled framework like change point models  to analyze  such unstationary data.


\begin{figure}[H]
    \centering
    \includegraphics[width = 0.32 \textwidth]{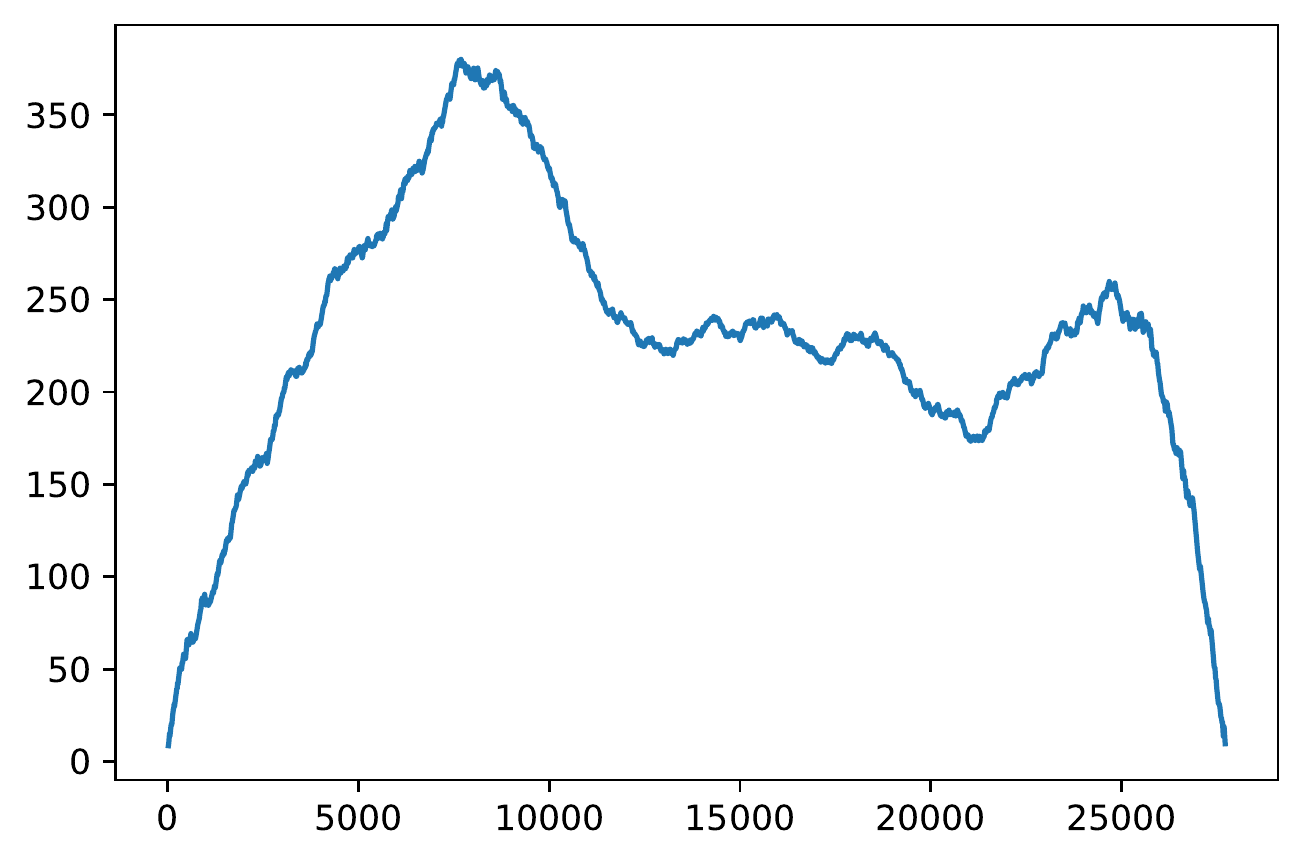}
    \includegraphics[width = 0.32 \textwidth]{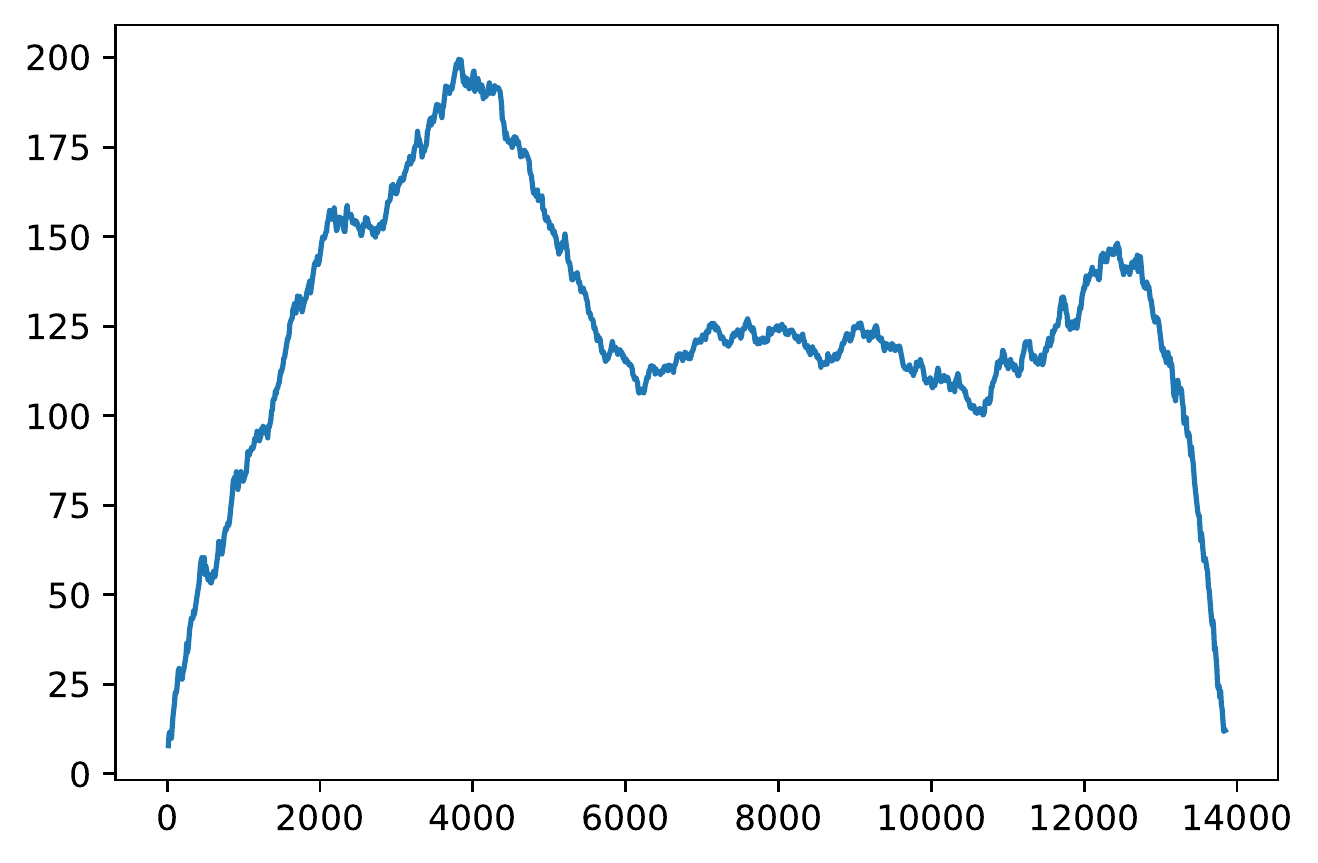}
    \includegraphics[width = 0.32 \textwidth]{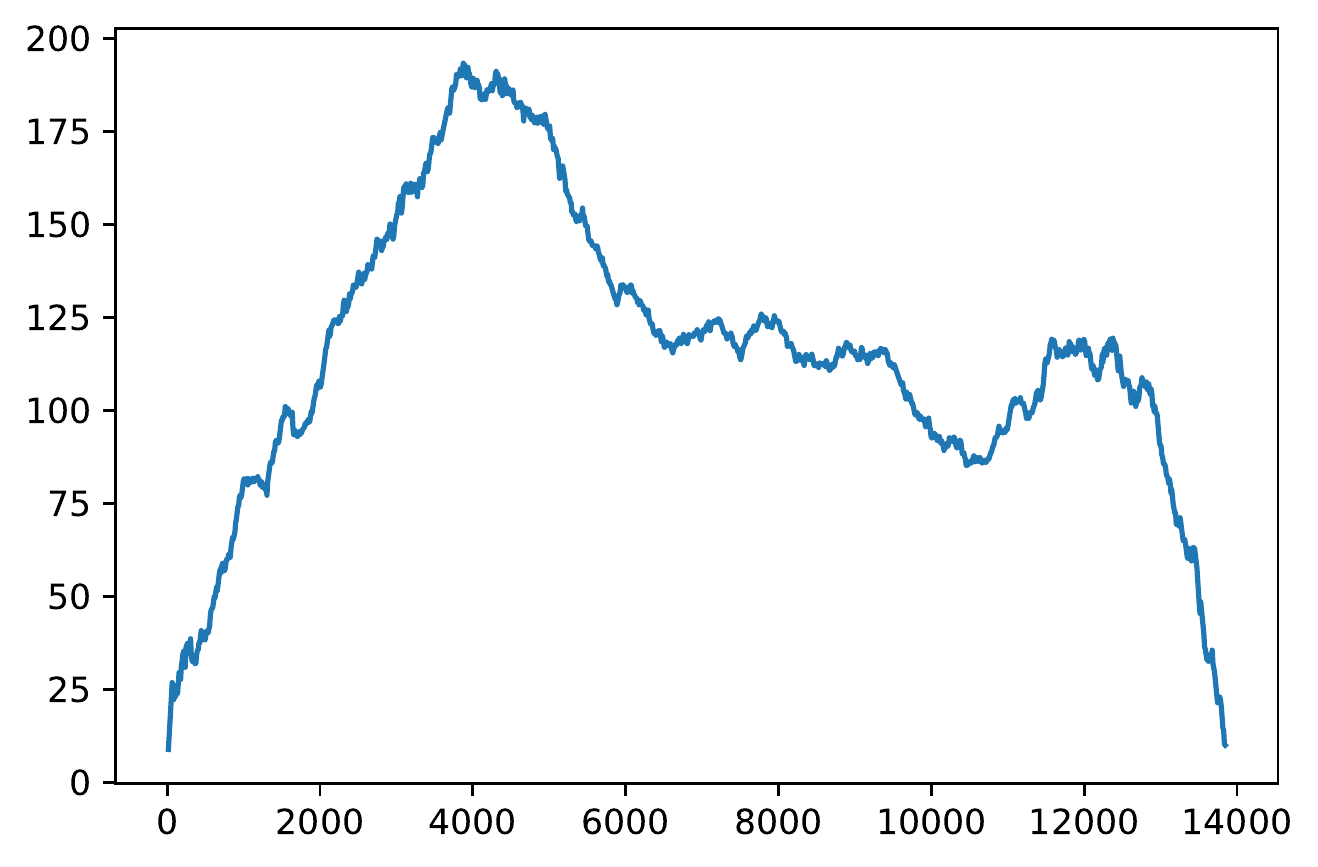}
    \caption{Path of (logarithmic) generalized loglikelihood ratio on NBA data. Left: GLR path on all samples; mid: GLR path on samples with odd indices; right: GLR path on samples with even indices.}
    \label{fig:nba_glr_2}
\end{figure}

To get a rough sense of the number and locations of change points, we check the paths  of the logarithm of generalized likelihood ratio statistics, which are shown in \Cref{fig:nba_glr_2}. It should be noted that although  the GLR paths suggest  the existence and locations of two change points, we cannot rely on these observation. This is because when multiple change points exist, there will be  cancellations effects and the GLR paths   may not give consistent estimates of change points \citep{Venkatraman1992}. We can also see that splitting the data by odd and even indices does not affect the shape of the     GLR path.   

With all the information in the exploratory analysis, we apply our method DPLR to the dataset and summarize results in \Cref{sec:application}.

\subsubsection{Comparison with WBS-GLR}
In this subsection, we apply the potential competitor, the likelihood-based WBS method  (i.e.  WBS-GLR), to the NBA data.
For a fair comparison, we set the regularization tuning parameter $\gamma$ in  the penalized logistic regression to be  $  0.1$, as we did in \Cref{sec:application} for DPLR. However, as   mentioned in \Cref{sec:wbs-glr}, WBS has another tuning parameter $M$, the number of random intervals to perform binary segmentation. So  we apply WBS-GLR with  $M\in \{50,100,150,200,250\}$, and  the estimated  change points with corresponding  test errors (negative log-likelihoods) are summarized in \Cref{tab:wbs_different_M}. Here, we use samples with odd time indices as training data and even
time indices as test data. It can be seen from \Cref{tab:wbs_different_M} that the choice of  $M$ does not have a significant impact on change point estimation in this real data example.   Therefore   in what follows,   we   only discuss    the results of  WBS-GLR with $M = 200$.

\begin{table}[H]
\centering
\scalebox{1}{
\begin{tabular}{cccc}
\hline
$M$ & Change point index & Change point season & Test errors \\
\hline
50 & $[7728, 14628, 20700, 24564]$ & [S1990m, S1999, S2007, S2012]  & 1796.9 \\
100 & $[7728, 14628, 20700, 24564]$ &  [S1990m, S1999, S2007, S2012] & 1796.9 \\
150 & $[7728, 14628, 20700, 24564]$ &  [S1990m, S1999, S2007, S2012] & 1796.9 \\
200 & $[7728, 14352, 20700, 24564]$ & [S1990m, S1998m, S2007, S2012]   & 1793.2 \\
250 & $[7728, 14628, 20700, 24564]$ & [S1990m, S1999, S2007, S2012]  & 1796.9 \\
\hline
\end{tabular}
}
\caption{The estimated change points with testing loss of WBS-GLR on the NBA data. S1980 means season 1980-1981, and S1990m means the middle of season 1990-1991}
\label{tab:wbs_different_M}

\end{table}



Then similar to \Cref{sec:application}, we    fit a  \btl{} model to  each interval segmented  by WBS-GLR, and summarize the results in \Cref{tab:nba_wbs}. As we can see, WBS-GLR is able to detect several important change points in the NBA history, e.g., the dominance of Celtics and Lakers in 1980s, the Bulls dynasty in 1990s, and the rise of Spurs afterwards. However, compared  with DPLR, WBS-GLR fails   to detect  the rise of Heat and Warriors. Therefore, the outcome of DPLR is more informative   in this real application, which again confirms our  findings in the simulation study in \Cref{sec:experiment}.
 
\begin{table}[H]
\centering
\setlength{\tabcolsep}{0.3em}
\scalebox{0.9}{\renewcommand{\arraystretch}{1.2}
\begin{tabular}{lrlrlrlrlr}
 \hline
\hline
\multicolumn{2}{c}{\textbf{S1980-S1990m}} & \multicolumn{2}{c}{\textbf{S1990m-S1998m}} & \multicolumn{2}{c}{\textbf{S1998m-S2006}} & \multicolumn{2}{c}{\textbf{S2007-S2011}} & \multicolumn{2}{c}{\textbf{S2012-S2015}} \\
\hline
Celtics                & 1.1137          & Bulls                  & 0.9435          & Spurs                  & 0.904           & Lakers                 & 0.7579          & Spurs                  & 1.1659          \\
Lakers                 & 1.084           & Jazz                   & 0.7996          & Mavericks              & 0.665           & Spurs                  & 0.701           & Clippers               & 0.9448          \\
76ers                  & 0.8049          & Suns                   & 0.5405          & Lakers                 & 0.5904          & Celtics                & 0.6406          & Warriors               & 0.9106          \\
Bucks                  & 0.7336          & Knicks                 & 0.5178          & Kings                  & 0.5103          & Magic                  & 0.6084          & Heat                   & 0.5149          \\
Pistons                & 0.5074          & Rockets                & 0.508           & Suns                   & 0.3677          & Mavericks              & 0.605           & Rockets                & 0.4703          \\
Trail Blazers          & 0.4466          & Trail Blazers          & 0.4931          & Timberwolves           & 0.2767          & Nuggets                & 0.458           & Mavericks              & 0.3402          \\
Suns                   & 0.284           & Spurs                  & 0.4638          & Pistons                & 0.2464          & Bulls                  & 0.2974          & Pacers                 & 0.3368          \\
Nuggets                & 0.2294          & Cavaliers              & 0.3415          & Jazz                   & 0.2266          & Suns                   & 0.28            & Trail Blazers          & 0.2782          \\
Bulls                  & 0.1782          & Lakers                 & 0.3338          & Pacers                 & 0.1902          & Rockets                & 0.2724          & Bulls                  & 0.2639          \\
Jazz                   & 0.1774          & Pacers                 & 0.241           & Rockets                & 0.0024          & Jazz                   & 0.2499          & Nuggets                & 0.0401          \\
Spurs                  & 0.1394          & Magic                  & 0.1824          & Trail Blazers          & -0.0049         & Trail Blazers          & 0.1843          & Jazz                   & -0.0495         \\
Rockets                & 0.1252          & Hornets                & 0.0923          & Heat                   & -0.0433         & Cavaliers              & 0.1628          & Cavaliers              & -0.0752         \\
Mavericks              & 0.1004          & Heat                   & 0.0572          & 76ers                  & -0.0673         & Hornets                & 0.0931          & Celtics                & -0.1486         \\
Knicks                 & 0.0744          & Pistons                & -0.1381         & Nets                   & -0.0807         & Heat                   & 0.081           & Hornets                & -0.1522         \\
Warriors               & -0.1406         & Warriors               & -0.2101         & Hornets                & -0.113          & 76ers                  & -0.157          & Nets                   & -0.2055         \\
Nets                   & -0.1751         & Celtics                & -0.2326         & Bucks                  & -0.2183         & Pistons                & -0.2651         & Knicks                 & -0.2865         \\
Pacers                 & -0.1857         & Nets                   & -0.3088         & Nuggets                & -0.2676         & Warriors               & -0.3028         & Suns                   & -0.296          \\
Cavaliers              & -0.2179         & Bucks                  & -0.473          & Magic                  & -0.2993         & Pacers                 & -0.3475         & Bucks                  & -0.354          \\
Kings                  & -0.3197         & Clippers               & -0.5024         & Knicks                 & -0.3218         & Bucks                  & -0.4778         & Pistons                & -0.3591         \\
Clippers               & -0.6276         & Kings                  & -0.5103         & Celtics                & -0.3293         & Knicks                 & -0.6236         & Kings                  & -0.4707         \\
Timberwolves           & -0.9485         & Nuggets                & -0.6578         & Clippers               & -0.4028         & Clippers               & -0.6919         & Lakers                 & -0.5136         \\
Hornets                & -1.0599         & Timberwolves           & -0.6859         & Cavaliers              & -0.4321         & Kings                  & -0.7288         & Timberwolves           & -0.5649         \\
Magic                  & -1.1178         & 76ers                  & -0.7395         & Warriors               & -0.5857         & Timberwolves           & -0.8974         & Magic                  & -0.697          \\
Heat                   & -1.206          & Mavericks              & -1.056          & Bulls                  & -0.8137         & Nets                   & -0.8998         & 76ers                  & -1.093 \\     
\hline
\end{tabular}
}
\caption{Fitted $\hat{\bbrtheta}$ (rounded to the fourth decimal) for 24 selected teams in seasons 1980-2016 of the National Basketball Association. Teams are ranked by the MLE $\hat{\bbrtheta}$ on subsets splitted at the estimated change points given by the WBS-GLR method. S1980 means season 1980-1981, and S1990m means the middle of season 1990-1991.}\label{tab:nba_wbs}
\end{table}

\subsection{Other potential competitors}
\label{sec: other competitor}

As we emphasized in \Cref{sec:introduction} and \Cref{sec:experiment}, localizing potential change points in pairwise comparison data is an unsolved problem. Given the good performance of our proposed method DPLR in this paper, one might wonder if there exist other methods that perform well, or even better than DPLR, in some aspects. This section intends to present some of our explorations on two potential efficient methods, WBS-SST and WBS-Mean. 

In what follows, we will demonstrate that both of them have crucial drawbacks. Specifically, WBS-Mean is not guaranteed to work for general comparison graphs, and works for general ranking models only under some constraints. WBS-SST works for general comparison graphs and ranking models, but requires relatively large sample size (i.e., $\Delta$) to work. Precise quantification of their performance can be an interesting direction for future works.

\subsubsection{Based on the test statistic for SST class}
\cite{nihar_siva2020} consider the two sample testing problem for general pairwise comparison data. Suppose we observe pairwise comparison outcome matrices $\mathbf{X}$ and $\mathbf{Y}$ generated from two winning probability matrices $\bfP, \bfQ\in \mathbb{R}^{n\times n}$, respectively. They propose the following test statistic:
\begin{equation}
\label{eq: SST two sample test}
    R_{SST}=\sum_{i=1}^d \sum_{j=1}^d \mathbb{I}_{i j} \frac{k_{i j}^q\left(k_{i j}^q-1\right)\left(X_{i j}^2-X_{i j}\right)+k_{i j}^p\left(k_{i j}^p-1\right)\left(Y_{i j}^2-Y_{i j}\right)-2\left(k_{i j}^p-1\right)\left(k_{i j}^q-1\right) X_{i j} Y_{i j}}{\left(k_{i j}^p-1\right)\left(k_{i j}^q-1\right)\left(k_{i j}^p+k_{i j}^q\right)},
\end{equation}
where $\mathbb{I}_{i j}=\mathbb{I}\left(k_{i j}^p>1\right) \times \mathbb{I}\left(k_{i j}^q>1\right)$, $k_{ij}^p = X_{ij} + X_{ji}$ and $k_{ij}^q = Y_{ij} + Y_{ji}$ are the number of comparisons between pairs.

We can use this test statistic to construct the loss $\mclR(t;s,e)$ in WBS (\Cref{algorithm:WBS}), i.e.,
\begin{equation}
    \mclR(t;s,e) = R_{SST}(\bfX([s,t)),\bfY([t,e))),
\end{equation}
and call this method WBS-SST (SST stands for strong stochastic transitive). When $\Delta$ is sufficiently large, WBS-SST performs fairly well with small computational cost, as is shown in \Cref{tab:compare_more_methods}.

\begin{figure}[H]
    \centering
    \includegraphics[width = 0.41 \textwidth]{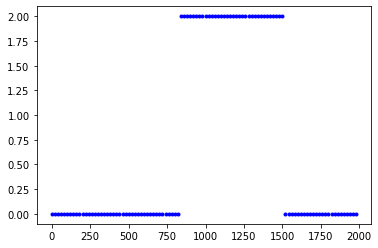}
    \includegraphics[width = 0.4 \textwidth]{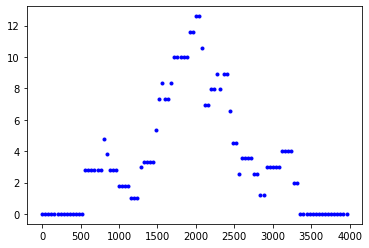}
    \caption{Loss path for WBS-SST when the sample size is not large enough. $n = 100$ with a single change point at the middle. $\Delta = 1000$ (left), $\Delta = 2000$ (right).}
    \label{fig: SST small Delta}
\end{figure}

\textbf{Issue with this approach.} However, When $\Delta$ is small, then many pairs in sampled intervals in WBS will have $k_{ij}\leq 1$, and the statistic would not be very powerful. See \Cref{fig: SST small Delta} and \Cref{tab:compare_more_methods}.

To see that reason, notice that
\begin{equation}
\begin{split}
        \mathbb{E}[R_{SST}|\mathbf{k}^p,\mathbf{k}^q] &= \sum_{i=1}^d \sum_{j=1}^d \mathbb{I}_{i j}\frac{k_{i j}^q\left(k_{i j}^q-1\right)k_{i j}^p\left(k_{i j}^p-1\right)(P_{ij}^2 +  Q_{ij}^2 - 2P_{ij}Q_{ij})}{\left(k_{i j}^p-1\right)\left(k_{i j}^q-1\right)\left(k_{i j}^p+k_{i j}^q\right)}\\
        &= \sum_{i=1}^d \sum_{j=1}^d \mathbb{I}_{i j}\frac{k_{i j}^qk_{i j}^p}{k_{i j}^p+k_{i j}^q}(P_{ij} -  Q_{ij})^2.
\end{split}
\end{equation}
When the comparison graph is a complete graph and compared pairs $\{(i_t,j_t)\}_{t\in [T]}$ are i.i.d. samples from the edge set $E_{full}:=\{(i,j):1\leq i< j\leq n\}$, the expectation of $R$ is (without the loss of generality, assume that $(1,2)\in E_{full}$)
\begin{equation}
    \mathbb{E}[R_{SST}] = \|P - Q\|_F^2\mathbb{E}[\mathbb{I}_{1,2}\frac{k_{1,2}^qk_{1,2}^p}{k_{1,2}^p+k_{1,2}^q}].
\end{equation}
The two equations above illustrate why WBS-SST does nor perform well in small-SNR cases.

\subsubsection{Based on the Borda count}
Borda count is a popular method in practice for ranking, due to its efficiency and generality \citep{shah2017simplerobustrankpairwisecomps}. Given an interval $\mclI$, the normalized Borda count vector is defined as
\begin{equation}
    \beta(\mathcal{I})_i = \frac{1}{|\mclI|}[N_w(i;\mclI) - N_l(i;\mclI)], \forall i\in [n],
\end{equation}
where $N_w(i;\mclI)$ and $N_l(i;\mclI)$ are the number of wining and loss of item $i$ in comparisons over the interval $\mclI$.

Since it is well-known in ranking literature that Borda count is not guaranteed to give consistent ranking for general comparison graphs, we only consider the complete graph here. When the comparison graph is a complete graph and compared pairs are i.i.d. samples from the edge set, and there is no change point in $\mclI$, the expectation of $\beta(\mathcal{I})_i$ is
\begin{equation}
\label{eq: expect borda}
    \mathbb{E}[\beta(\mathcal{I})_i] = \frac{2}{n(n-1)}\sum_{j\neq i}(P_{ij} - P_{ji}) = \frac{2}{n(n-1)}\sum_{j\neq i}(2P_{ij} - 1),
\end{equation}
where $P_{ij} = \mathbb{P}[i \text{ beats } j]$.

If we treat $\beta(\mathcal{I})$ as a sample mean of a random variable, we can construct the CUSUM statistic at $t\in \mclI = [s, e)$ as
\begin{equation}
    \mathcal{R}_{Borda}(t;[s,e)) = \frac{(t - s )(e-t)}{e - s}\|\bbrbeta([s,t))-\bbrbeta([t,e))\|_2^2.
\end{equation} 
To compared this statistic with \Cref{eq: SST two sample test}, we assume there is a single change point $\eta\in [s,e)$ and check the statistic at $\eta$. By \Cref{eq: expect borda}, the population version of the statistic is
\begin{equation}
 \begin{split}
      \widetilde{\mathcal{R}}_{Borda}(\eta;[s,e)) & =\frac{(\eta - s )(e-\eta)}{e - s}\|\mathbb{E}\bbrbeta([s,\eta))-\mathbb{E}\bbrbeta([\eta,e))\|_2^2 \\
      &= \frac{(\eta - s )(e-\eta)}{e - s}\cdot \frac{2}{n(n-1)}\sum_{i\in [n]}[\sum_{j\neq i} (P_{ij} - Q_{ij})]^2
 \end{split}
 \label{eq: population_R_wbs-mean}
\end{equation}
where $\bfP,\bfQ$ are the winning probability matrices before and after the change point $\eta$.

\begin{figure}[H]
    \centering
    \includegraphics[width = 0.41 \textwidth]{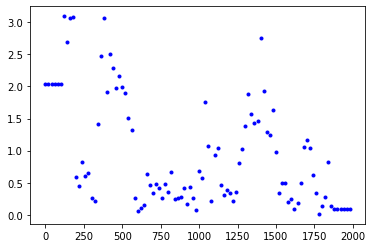}
    \includegraphics[width = 0.4 \textwidth]{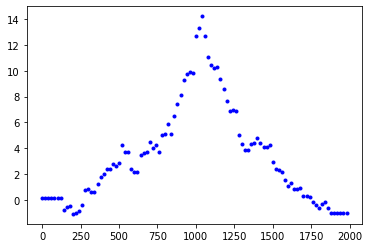}
    \caption{Loss path for WBS-Mean (left) and WBS-SST (right).}
    \label{fig: mean SST counter example}
\end{figure}

\paragraph{Issue with this approach.} With \Cref{eq: population_R_wbs-mean}, we can construct examples such that the population version of the CUSUM statistic is very small or even zero at the true change point $\eta$. For instance, let $n = 3$ and
\begin{equation}
    \bfP = \left[\begin{array}{ccc}
        0.5 & 0.6 & 0.8 \\
        0.4 & 0.5 & 0.7 \\
        0.2 & 0.3 & 0.5
    \end{array}
    \right],\quad 
    \bfQ = \left[\begin{array}{ccc}
        0.5 & 0.55 & 0.85 \\
        0.45 & 0.5 & 0.65 \\
        0.15 & 0.35 & 0.5
    \end{array}
    \right],
\end{equation}
then both $\bfP, \bfQ$ are strong-stochastic-transitive matrices (see \cite{shah2017simplerobustrankpairwisecomps} for details) and the population CUSUM $\widetilde{\mathcal{R}}_{Borda}(\eta;[s,e)) = 0$ at $\eta$. \Cref{fig: mean SST counter example} compares paths of the loss for WBS-Mean and WBS-SST under the choice of $\bfP,\bfQ$ above, where there is a single change point at $\eta = 1000$.

\subsubsection{Numerical performance}
\Cref{tab:compare_more_methods} compares the performance of WBS-SST and WBS-Mean with the two methods presented in the main text, under the identical setting in \Cref{sec:experiment}. The setting is sketched below for convenience.

\begin{table}[!h]
 \centering
 \begin{tabular}{m{2cm}m{2cm}m{2cm}C{1.5cm}C{1.5cm}C{1.5cm}}
  \hline
  & $H(\hat{\eta},\eta)$ & Time & $\hat{K}<K$ &$\hat{K}=K$ &$\hat{K}>K$
 \\
 \hline
 \multicolumn{6}{c}{{\bf Setting (i)} \ \ $n = 10, K = 3, \Delta = 500$, Change (I, II, III)} \\
 DPLR & 9.2 (9.1) & 49.7s (0.7) & 0 & 100 & 0 \\
 WBS-Mean & 15.4 (8.4) & 0.2s (0.05) & 0 & 100 & 0 \\
 WBS-SST & 16.2 (11.4) & 0.4s (0.2) & 0 & 100 & 0 \\
 WBS-GLR & 15.2 (7.9) & 31.9s (3.9) & 0 & 100 & 0 \\
 \hline
 \multicolumn{6}{c}{ {\bf Setting (ii)} \ \ $n = 20, K = 3, \Delta = 800$, Change (I, II, III)} \\
 DPLR & 9.0 (9.9) & 118.5s (2.2) & 0 & 100 & 0 \\
 WBS-Mean & 5.8 (11.4) & 0.5s (0.1) & 0 & 100 & 0 \\
 WBS-SST & 19.4 (22.3) & 1.7s (0.5) & 0 & 100 & 0 \\
 WBS-GLR & 240.5 (220.3) & 144.2s (12.5) & 0 & 40 & 60 \\
 \hline
 \multicolumn{6}{c}{{\bf Setting (iii)}\ \ $n = 100, K = 2, \Delta = 1000$, Change (I, II)} \\
 DPLR & 13.4 (14.4) & 167.4s (3.3) & 0 & 100 & 0 \\
 WBS-Mean & 22.9 (98.4) & 0.6s (0.04) & 1 & 99 & 0 \\
 WBS-SST & $\infty$ (NA) & 3.9s (0.4) & 100 & 0 & 0 \\
 WBS-GLR & 111.9 (195.6) & 215.9s (17.0) & 0 & 79 & 21 \\
 \hline
 \multicolumn{6}{c}{{\bf Setting (iv)} \ \ $n = 100, K = 3, \Delta = 2000$, Change (I, II, III)} \\
 DPLR & 12.4 (12.1) & 402.4s (7.4) & 0 & 100 & 0 \\
 WBS-Mean & 17.9 (6.1) & 0.9s (0.06) & 0 & 100 & 0 \\
 WBS-SST & 1116.3 (694.8) & 19.3s (1.9) & 57 & 42 & 1 \\
 WBS-GLR & 412.3 (495.5) & 400.0s (40.9) & 0 & 57 & 43 \\
 \hline
 \end{tabular}
 \caption{Comparison of DPLR and three WBS-based methods under four different simulation settings. 100 trials are conducted in each setting. For the localization error and running time (in seconds), the average over 100 trials is shown with standard error in the bracket. The three columns on the right record the number of trials in which $\hat{K}<K$, $\hat{K}=K$, and $\hat{K}>K$ respectively.}
 \label{tab:compare_more_methods}
\end{table}

\paragraph{Simulation settings.} 
For $1< i \le n$, we set $\theta_i^*(\eta_0) = \theta_1^*(\eta_0) +(i -1)\delta$ with some constant $\delta$. In each experiment, we set $\delta$ first and then set $\theta_1^*(\eta_0)$ to make $\mathbf{1}^\top_n\bbrtheta^*(\eta_0) = 0$. The value of $\delta$ guarantees that the maximum winning probability is 0.9. We consider three types of changes:

Type I (reverse): $\theta_i^*(\eta_k) = \theta^*_{n + 1 - i}(\eta_0)$.

Type II (block-reverse): $\theta_i^*(\eta_k) = \theta^*_{[\frac{n}{2}] + 1 - i}(\eta_0)$ for $i\leq [\frac{n}{2}]$; $\theta_i^*(\eta_k) = \theta^*_{[\frac{n}{2}] + n + 1 - i}(\eta_0)$ for $i> [\frac{n}{2}]$.

Type III (block exchange): $\theta_i^*(\eta_k) = \theta^*_{i + [\frac{n}{2}]}(\eta_0)$ for $i\leq [\frac{n}{2}]$; $\theta_i^*(\eta_k) = \theta^*_{i - [\frac{n}{2}]}(\eta_0)$ for $i> [\frac{n}{2}]$.

We consider four simulation settings. For each setting, we have $T = (K+1)\Delta$ and the change points locate at $\eta_i = i \Delta$ for $i\in [K]$. To describe the true parameter at each change point, we use an ordered tuple. For instance, (I, II, III, I) means that $K=4$ and the true parameters at $\eta_1,\eta_2,\eta_3, \eta_4$ are determined based on $\bbrtheta^*(\eta_0)$ and the change type I, II, III, and I, respectively.

\clearpage
\section{Appendix: Proof}
\label{sec:appendix}
This section has three parts:
\begin{enumerate}
    \item \Cref{sec:proof_thm} contains the proof of two theorems in \Cref{sec:main_result}.
    \item \Cref{sec:props} contains propositions used throughout the paper with proof.
    \item \Cref{sec:lemmas} contains all technical lemmas with proof.
\end{enumerate}

\subsection{Proof of main theorems}
\label{sec:proof_thm}
\bprf[Proof of \Cref{thm:consistency_bt}]
The theorem is a straightforward conclusion of \Cref{prop:4case_bt} and \Cref{prop:P_K}. More specifically, conclusion 3 and 4 of \Cref{prop:4case_bt} guarantee that $K\leq |\hat{\mclP}|\leq 3K$ with probability at least $1 - (Tn)^{-2}$ and \Cref{prop:P_K} further confirms the consistency of $\hat{K}$. Then conclusion 1 and 2 of \Cref{prop:4case_bt} control the localization error.
\eprf

\bprf[Proof of \Cref{thm:rate_dp_local_refine}]
The theorem is a straightforward conclusion of \Cref{thm:consistency_bt} that quantifies the localization error of outputs of dynamic programming and \Cref{prop:local_refine} that shows the improvement of local refinement.
\eprf

\subsection{Main propositions}
\label{sec:props}

\bnprop[Consistency of $\hat{K}$]
\label{prop:P_K}
Let $\hat{\mclP}$ be the estimator of change points in \Cref{eq:P_hat_bt}. Assume $K\leq |\hat{\mclP}|\leq 3K$. Under all assumptions above, it holds with probability at least $1 - (Tn)^{-2}$ that $|\hat{\mclP}| = K$.
\enprop
\bprf
For a sequence of strictly increasing integer time points $\{\eta_j'\}_{j\in [J+1]}$ with $\eta_0' = 1$ and $\eta_{J+1}' = T + 1$, let $\mclI_j = [\eta_{j-1}',\eta_{j}')$ and
\begin{equation*}
    L(\{\eta_j'\}_{j\in [J+1]}) = \sum_{j\in [J + 1]} L(\hat{\bbrtheta}(\mclI_j),\mclI_j),
\end{equation*}
where $\hat{\bbrtheta}(\mclI_j):=\argmin_{\bbrtheta\in \Theta_B}L({\bbrtheta},\mclI_j)$. Furthermore, when $\{\eta_k\}_{k\in [K]}\subset \{\eta_j'\}_{j\in [J+1]}$ so that $\theta^*(t)$ remains unchanged in each interval $\mclI_j$, we can define the risk of true parameters
\begin{equation*}
    L^*(\{\eta_j'\}_{j\in [J+1]}) = \sum_{j\in [J + 1]} L({\bbrtheta}^*(\mclI_j),\mclI_j).
\end{equation*}
Let $\{\hat{\eta}_k\}_{k\in [\hat{K}]}$ be the change points given by the estimator $\hat{\mclP}$ and $\mathbf{Sort}(\cdot)$ be an operator on finite ordered tuple of scalars such that $\mathbf{Sort}((a_1,\ldots,a_m)) = (a_{(1)},\ldots,a_{(m)})$ where $a_{(i)}\leq a_{(j)}$ for $i< j$ and $\{a_{(i)}\}_{i\in [m]} = \{a_{i}\}_{i\in [m]}$. Then a sufficient condition for $|\hat{\mclP}|=K$ is
\begin{align}
    &L^*(\eta_1,\cdots,\eta_K) + K\gamma \nonumber \\
    \geq & L(\eta_1,\cdots,\eta_K) + K\gamma \label{eq:prop2_c1_e1_bt}\\ 
    \geq & L(\hat{\eta}_1,\cdots,\hat{\eta}_{\hat{K}}) + \hat{K}\gamma \label{eq:prop2_c1_e2_bt}\\ 
    \geq & L^*(\mathbf{Sort}(\hat{\eta}_1,\cdots,\hat{\eta}_{\hat{K}},\eta_1,\cdots,\eta_K)) + \hat{K}\gamma - CK p_{lb}^{-2}\frac{nd_{\max}}{\lambda_2(L_{\mclG})}\log(Tn), \label{eq:prop2_c1_e3_bt}
\end{align}
and
\begin{equation}
    L^*(\eta_1,\cdots,\eta_K)\leq L^*(\mathbf{Sort}(\hat{\eta}_1,\cdots,\hat{\eta}_{\hat{K}},\eta_1,\cdots,\eta_K)) + C_1 K p_{lb}^{-2}\frac{nd_{\max}}{\lambda_2(L_{\mclG})}\log(Tn).
    \label{eq:prop2_c2_bt}
\end{equation}
In fact, if $\hat{K}\geq K+1$, under the conditions above and the assumption that $|\hat{\mclP}|\leq 3K$, we have
\begin{equation*}
    \gamma\leq (\hat{K} - K)\gamma \leq  C_2 Kp_{lb}^{-2} \frac{nd_{\max}}{\lambda_2(L_{\mclG})} \log(Tn),
\end{equation*}
which is contradictory to the definition $\gamma = C_{\gamma}(K+1)p_{lb}^{-2}\frac{nd_{\max}}{\lambda_2(L_{\mclG})}\log (Tn)$ for sufficiently large $C_{\gamma}$.

Now we prove that the sufficient condition holds with probability at least $1 - (Tn)^{-2}$. \Cref{eq:prop2_c1_e1_bt} is a straightforward conclusion of the definition $\hat{\bbrtheta}(\mclI_j):=\argmin_{\bbrtheta\in \Theta_B}L({\bbrtheta},\mclI_j)$ and \Cref{eq:prop2_c1_e2_bt} is implied by the definition of $\hat{\mclP}$ in \Cref{eq:P_hat_bt}.

\Cref{eq:prop2_c2_bt} is guaranteed by \Cref{lem:l_hat_lower_bt} because for any interval $\mclI$ determined by endpoints that are two consecutive points in $\mathbf{Sort}(\hat{\eta}_1,\cdots,\hat{\eta}_{\hat{K}},\eta_1,\cdots,\eta_K)$, there will not be any true change point in the interior of $\mclI$.

For \Cref{eq:prop2_c1_e3_bt}, notice that by \Cref{prop:4case_bt}, with probability $1 - (Tn)^{-4}$, there are at most two change points in $\mclI$. Therefore, \Cref{lem:3case_bt} ensures that
\begin{align*}
        L(\hat{\eta}_1,\cdots,\hat{\eta}_{\hat{K}}) \geq L^*(\mathbf{Sort}(\hat{\eta}_1,\cdots,\hat{\eta}_{\hat{K}},\eta_1,\cdots,\eta_K)) - CKp_{lb}^{-2}\frac{nd_{\max}}{\lambda_2(L_{\mclG})}\log (Tn).
\end{align*}
\eprf

\bnprop[Four cases]
\label{prop:4case_bt}
Let $\hat{\mclP}$ be the estimator of change points in \Cref{eq:P_hat_bt}. Under \cref{assp:snr_bt} and \Cref{assp:glm_bt}, with probability at least $1 - (Tn)^{-2}$ the following four events hold uniformly for all $\mclI = (s,e)\in \hat{\mclP}$:
\begin{enumerate}
    \item If $\mclI$ contains only one change point $\eta$, then for some universal constant $C>0$,
    \begin{equation*}
        \min\{\eta - s,e-\eta\}\leq Cp_{lb}^{-2}\frac{|E|}{\lambda_2(L_{\mclG})}[\gamma + \frac{nd_{\max}}{\lambda_2(L_{\mclG})}\log(Tn)].
    \end{equation*}
    \item If $\mclI$ contains exactly two change points $\eta_k$ and $\eta_{k+1}$, then for some universal constant $C>0$,
    \begin{equation*}
        \min\{\eta_{k} - s,e-\eta_{k+1}\}\leq Cp_{lb}^{-2}\frac{|E|}{\lambda_2(L_{\mclG})}[\gamma + \frac{nd_{\max}}{\lambda_2(L_{\mclG})}\log(Tn)].
    \end{equation*}
    \item If $|\hat{\mclP}|>1$, then for any two consecutive intervals $\mclI$ and $\mclJ$ in $\hat{\mclP}$, the joint interval $\mclI\cup \mclJ$ contains at least one change point.
    \item Interval $\mclI$ does not contain more than two change points.
\end{enumerate}
\enprop
\bprf
Conclusion 1 is implied by \Cref{lem:1_change} and conclusion 2 is guaranteed by \Cref{lem:2_change}. Conclusion 4 is a direct consequence of \Cref{lem:3more_change} and the definition of $\hat{\mclP}$.

To prove conclusion 3, assume instead that there is no true change point in $\mclI\cup \mclJ$. Then by \Cref{lem:no_change} we have
\begin{equation*}
    L(\hat{\bbrtheta}(\mclI),\mclI) + L(\hat{\bbrtheta}(\mclJ),\mclJ) + \gamma \geq L({\bbrtheta}^*(\mclI\cup \mclJ),\mclI\cup \mclJ)\geq L(\hat{\bbrtheta}(\mclI\cup \mclJ),\mclI\cup \mclJ),
\end{equation*}
which is contradictory to the definition of $\hat{\mclP}$.
\eprf

\bnprop[Local refinement]
\label{prop:local_refine}
Consider the local refinement procedure given in \Cref{algo:local_refine_bt}, that is,
\begin{align}
    \left(\hat{\eta}_k,\hat{\bbrtheta}^{(1)},\hat{\bbrtheta}^{(2)}\right) =\argmin_{\substack{\eta \in \{s_k + 1, \ldots, e_k - 1\} \\ \bbrtheta^{(1)}, \bbrtheta^{(2)} \in \Theta_{B}}} \Bigg\{\sum_{t = s_k + 1}^{\eta}\ell_t(\bbrtheta^{(1)}) + \sum_{t = \eta + 1}^{e_k}\ell_t(\bbrtheta^{(2)})\Bigg\},
\end{align}
where $s_k=2\widetilde{\eta}_{k-1}/3 + \widetilde{\eta}_{k}/3$ and $ e_k =  \widetilde{\eta}_{k}/3 + 2\widetilde{\eta}_{k+1}/3$ and $\ell_t(\bbrtheta)$ is the negative log-likelihood given in \Cref{eq:likelihood}. Suppose the input $\{\tilde{\eta}_{k}\}_{k\in [\tilde{K}]}$ satisfies $\tilde{K} = K$ and
\begin{equation*}
    \max_{k\in [K]}|\tilde{\eta}_k - \eta_k|\leq \Delta / 5.
\end{equation*}
Let $\{\hat{\eta}_k\}_{k \in [K]}$ be the output. Then it holds with probability at least $1 - (Tn)^{-2}$ that
\begin{equation}
    \max_{k\in [K]}|\hat{\eta}_k - \eta_k|\leq C\frac{|E|nd_{\max}}{p_{lb}^{2}\kappa^2\lambda_2^2(L_{\mclG})}\log(Tn).
\end{equation}
\enprop

\bprf
For each $k\in [K]$, let $\hat{\bbrtheta}(t) = \hat{\bbrtheta}^{(1)}$ if $s_k<t\leq \hat{\eta}_k$ and $\hat{\bbrtheta}(t) =\hat{\bbrtheta}^{(2)}$ otherwise, and ${\bbrtheta}^*(t)$ be the true parameter at time point $t$. First we show that under conditions $\tilde{K} = K$ and $\max_{k\in [K]}|\tilde{\eta}_k - \eta_k|\leq \Delta / 5$, there is only one true change point $\eta_k$ in $(s_k, e_k)$. It suffices to show that
\begin{equation}
    |\tilde{\eta}_k - \eta_k|\leq \frac{2}{3}(\tilde{\eta}_{k+1} - \tilde{\eta}_{k}),\ \text{and}\ |\tilde{\eta}_{k+1} - \eta_{k+1}|\leq \frac{1}{3}(\tilde{\eta}_{k+1} - \tilde{\eta}_{k}).
    \label{tmp_eq:goal_lr}
\end{equation}
Denote $R = \max_{k\in [K]}|\tilde{\eta}_k - \eta_k|$, then
\begin{align*}
    \tilde{\eta}_{k+1} - \tilde{\eta}_{k} &= \tilde{\eta}_{k+1} - {\eta}_{k+1} + {\eta}_{k+1} - {\eta}_{k} + {\eta}_{k} - \tilde{\eta}_{k} \\
    &= ({\eta}_{k+1} - {\eta}_{k}) + (\tilde{\eta}_{k+1} - {\eta}_{k+1}) + ({\eta}_{k} - \tilde{\eta}_{k})\in [{\eta}_{k+1} - {\eta}_{k} - 2R,{\eta}_{k+1} - {\eta}_{k} + 2R].
\end{align*}
Therefore, \Cref{tmp_eq:goal_lr} is guaranteed as long as
\begin{equation*}
    R\leq \frac{1}{3}(\Delta - 2R),
\end{equation*}
which is equivalent to $R\leq \Delta/5$.

Now without loss of generality, assume that $s_k<\eta_k<\hat{\eta}_k<e_k$. Denote $\mclI_k = \{s_k + 1, \cdots, e_k\}$. Consider two cases:

\textbf{Case 1} If
\begin{equation*}
    \hat{\eta}_k - \eta_k < \max\{Cp_{lb}^{-2}\frac{|E|nd_{\max}}{\lambda_2^2(L_{\mclG})}\log(Tn), Cp_{lb}^{-2}\log(Tn) / \kappa^2\},
\end{equation*}
then the proof is done.

\textbf{Case 2} If
\begin{equation*}
    \hat{\eta}_k - \eta_k \geq \max\{Cp_{lb}^{-2}\frac{|E|nd_{\max}}{\lambda_2^2(L_{\mclG})}\log(Tn), Cp_{lb}^{-2}\log(Tn) / \kappa^2\},
\end{equation*}
then we proceed to prove that $|\hat{\eta}_k - \eta_k|\leq C_1\frac{|E|nd_{\max}}{p_{lb}^{2}\kappa^2\lambda_2^2(L_{\mclG})}\log(Tn)$ with probability at least $1 - (Tn)^{-3}$. Then we either prove the result or get an contradiction, and complete the proof in either case.

By the definition of $\hat{\eta}_k,\hat{\bbrtheta}^{(1)}$, and $\hat{\bbrtheta}^{(2)}$, we have
\begin{equation*}
    \sum_{t\in \mclI_k}\ell_t(\hat{\bbrtheta}(t))\leq \sum_{t\in \mclI_k}\ell_t({\bbrtheta}^*(t)).
\end{equation*}
By \Cref{lem:convex_bt}, this implies that
\begin{equation}
    ce^{-2B}\sum_{t\in \mclI_k}[\bfx(t)^\top \Delta(t)]^2 \leq \sum_{t\in \mclI_k}\epsilon_t \bfx(t)^\top \Delta(t),
    \label{teq:first}
\end{equation}
where $\Delta(t) := \hat{\bbrtheta}(t) - \bbrtheta^*(t)$ and $\epsilon_t := y_t - \frac{\exp(\bfx(t)^\top \bbrtheta^*(t))}{1 + \exp(\bfx(t)^\top \bbrtheta^*(t))}$. For the cross term, by \Cref{lem:R_cp_bt} we have
\begin{align}
    \sum_{t\in \mclI_k}\epsilon_t \bfx(t)^\top \Delta(t) &=\sum_{i\in [n]}\{ \left|\frac{\sum_{t\in \mclI_k  }\epsilon_t x_i(t)\Delta_i(t)}{\sqrt{\sum_{t\in \mclI_k}\Delta_i(t)^2}}\right| \sqrt{\sum_{t\in \mclI_k}\Delta_i(t)^2}\} \nonumber\\
    &\leq \sup_{i\in [n]}\left|\frac{\sum_{t\in \mclI_k  }\epsilon_t x_i(t)\Delta_i(t)}{\sqrt{\sum_{t\in \mclI_k}\Delta_i(t)^2}}\right|\sum_{i\in [n]}\sqrt{\sum_{t\in \mclI_k}\Delta_i(t)^2}  \nonumber\\ 
    &\leq  C\sqrt{\frac{d_{\max}}{|E|}\log(Tn)} \sum_{i\in [n]}\sqrt{\sum_{t\in \mclI_k} (\hat{\theta}_i - \theta^*_i(t))^2}  \nonumber\\
        &\leq C\sqrt{\frac{nd_{\max}}{|E|}\log(Tn)}\sqrt{\sum_{t\in \mclI_k}\|\Delta(t)\|_2^2}. \label{teq:cross}
\end{align}
\Cref{teq:first} and \Cref{teq:cross} together imply that
\begin{equation}
    ce^{-2B}\sum_{t\in \mclI_k}[\bfx(t)^\top \Delta(t)]^2 \leq C\sqrt{\frac{nd_{\max}}{|E|}\log(Tn)} \sqrt{\sum_{t\in \mclI_k}\|\Delta(t)\|_2^2}.
\end{equation}
Let
\begin{equation*}
    \mclJ_1 = (s_k,\eta_k],\ \mclJ_2 = (\eta_k, \hat{\eta}_k],\ \mclJ_3 = (\hat{\eta}_k, e_k].
\end{equation*}
Under \Cref{assp:snr_bt} and the condition of the proposition, it holds that $\min\{|\mclJ_1|,|\mclJ_3|\}\geq C_0\frac{|E|\log(Tn)}{\lambda_2(L_{\mclG})}$. Thus, by \Cref{lem:eig_laplacian_gnm}, with probability at leat $1 - (Tn)^{-3}$, we have
\begin{align*}
    \sum_{t\in \mclI_k}[\bfx(t)^\top \Delta(t)]^2 \geq \frac{c_1\lambda_2(L_{\mclG})}{|E|}\sum_{t\in \mclI_k}\|\Delta(t)\|_2^2.
\end{align*}
The inequality above leads to
\begin{equation*}
    \sum_{t\in \mclI_k}\|\hat{\bbrtheta}(t) - \bbrtheta^*(t)\|_2^2 \leq Cp_{lb}^{-2} \frac{|E|nd_{\max}}{\lambda_2^2(L_{\mclG})} \log(Tn).
\end{equation*}
Recall that we defined $\bbrtheta^{(1)} = \bbrtheta^*({\eta_k -1})$ and $\bbrtheta^{(2)} = \bbrtheta^*({\eta_k})$. Then we have
\begin{equation*}
    \sum_{t\in \mclI_k} \|\hat{\bbrtheta}(t) - \bbrtheta^*(t)\|_2^2 = |\mclJ_1|\|\hat{\bbrtheta}^{(1)} - \bbrtheta^{(1)}\|_2^2 + |\mclJ_2|\|\hat{\bbrtheta}^{(1)} - \bbrtheta^{(2)}\|_2^2 + |\mclJ_3|\|\hat{\bbrtheta}^{(2)} - \bbrtheta^{(2)}\|_2^2.
\end{equation*}
Since $|\mclJ_1|=\eta_k - s_k\geq c_0\Delta$ with some constant $c_0$ under \Cref{assp:snr_bt}, we have
\begin{equation}
    \Delta \|\hat{\bbrtheta}^{(1)} - \bbrtheta^{(1)}\|_2^2\leq c_0|\mclJ_1|\|\hat{\bbrtheta}^{(1)} - \bbrtheta^{(1)}\|_2^2 \leq c_1p_{lb}^{-2}\frac{|E|nd_{\max}}{\lambda_2^2(L_{\mclG})}\log(Tn)\leq c_2\Delta \kappa^2,
\end{equation}
with some constant $c_2\in (0,1/4)$, where the last inequality is due to the fact that $\mathcal{B}_T\rightarrow \infty$. Thus we have
\begin{equation*}
    \|\hat{\bbrtheta}^{(1)} - \bbrtheta^{(1)}\|_2^2\leq c_2\kappa^2.
\end{equation*}
Triangle inequality gives
\begin{equation*}
    \|\hat{\bbrtheta}^{(1)} - \bbrtheta^{(2)}\|_2\geq \|\bbrtheta^{(1)} - \bbrtheta^{(2)}\|_2 - \|\hat{\bbrtheta}^{(1)} - \bbrtheta^{(1)}\|_2 \geq \kappa/2.
\end{equation*}
Therefore, $\kappa^2|\mclJ_2|/4\leq |\mclJ_2|\|\hat{\bbrtheta}^{(1)} - \bbrtheta^{(2)}\|_2^2\leq Cp_{lb}^{-2}\frac{|E|nd_{\max}}{\lambda_2^2(L_{\mclG})}\log(Tn)$ and
\begin{equation*}
    |\hat{\eta}_k - \eta_k| = |\mclJ_2|\leq \frac{Cp_{lb}^{-2}|E|nd_{\max}\log(Tn)}{\lambda_2^2(L_{\mclG})\kappa^2}.
\end{equation*}
\eprf

\bnprop
\label{prop:prob_matrix_theta}
Let $\bfP(\bbrtheta)$ be the winning probability matrix induced by $\bbrtheta$. For $\bbrtheta^{(1)},\bbrtheta^{(2)}\in \Theta_B$, it holds that
\begin{equation}
    \frac{np^2_{lb}}{16}\|\bbrtheta^{(1)} - \bbrtheta^{(2)}\|_2^2\leq \|\bfP(\bbrtheta^{(1)}) - \bfP(\bbrtheta^{(2)})\|_F^2\leq \frac{n}{16}\|\bbrtheta^{(1)} - \bbrtheta^{(2)}\|_2^2,
\end{equation}
where $p_{lb} = \frac{e^{-2B}}{1 + e^{-2B}}$.
\enprop
\bprf
This result has been shown in \cite{shah2017simplerobustrankpairwisecomps} (In the proof of Theorem 4). We include it here for completeness.

Denote $\psi(t) = \frac{1}{1 + e^{-t}}$. For any pair $(i,j)\in [n]^2$, by the mean value theorem we have
\begin{align*}
    |P_{ij}(\bbrtheta^{(1)}) - P_{ij}(\bbrtheta^{(2)})| &= |\psi(\theta^{(1)}_i - \theta^{(1)}_j) - \psi(\theta^{(2)}_i - \theta^{(2)}_j)|\\
    &= |\psi'(\xi)||(\theta^{(1)}_i - \theta^{(1)}_j) - (\theta^{(2)}_i - \theta^{(2)}_j)|,
\end{align*}
where $\xi$ is a scalar between $(\theta^{(1)}_i - \theta^{(1)}_j)$ and $(\theta^{(2)}_i - \theta^{(2)}_j)$. Since $\psi'(t) = \psi(t)(1 - \psi(t))\in (\frac{1}{4e^{2B}}, \frac{1}{4}]$ for $t\in [-2B,2B]$, we have
\begin{equation*}
    \frac{1}{4e^{2B}}|(\theta^{(1)}_i - \theta^{(1)}_j) - (\theta^{(2)}_i - \theta^{(2)}_j)|\leq |P_{ij}(\bbrtheta^{(1)}) - P_{ij}(\bbrtheta^{(2)})| \leq \frac{1}{4}|(\theta^{(1)}_i - \theta^{(1)}_j) - (\theta^{(2)}_i - \theta^{(2)}_j)|.
\end{equation*}
By the property of Graph Laplacian and the fact that $\mathbf{1}_n^\top \bbrtheta^{(i)} = 0$, $i = 1,2$, we have
\begin{align}
        \sum_{i,j\in [n]^2}[(\theta^{(1)}_i - \theta^{(1)}_j) - (\theta^{(2)}_i - \theta^{(2)}_j)]^2 &= (\bbrtheta^{(1)} - \bbrtheta^{(2)})^\top[n\bfI_{n} - \mathbf{1}_n\mathbf{1}_n^\top] (\bbrtheta^{(1)} - \bbrtheta^{(2)})\\
        &=n\|\bbrtheta^{(1)} - \bbrtheta^{(2)}\|_2^2.
\end{align}
Combining the results above gives the conclusion.
\eprf

\bnprop[Single change point]
\label{prop:single_cp}
Suppose we observe $\{(\bfx(t),y_t)\}_{t\in [T]}$ following model \eqref{eq:random_graph} and \eqref{eq:model_bt} and there is a single change point $\eta\in (1,T)$. In addition, assume that
\begin{equation}
    \Delta:= \min\{\eta - 1, T - \eta\}\geq \mathcal{B}_T\frac{|E|nd_{\max}}{p_{lb}^{2}\kappa^2\lambda_2^2(L_{\mclG})}\log(Tn),
    \label{eq:snr_single_cp}
\end{equation}
for a diverging sequence $\{\mathcal{B}_T\}_{T\in \mathbb{Z}_+}$. Let the estimator $\hat{\eta}$ of the change point be
\begin{align}
    \hat{\eta} =\argmin_{\eta \in [T]} \Bigg\{\min_{\bbrtheta^{(1)}\in \Theta_B}\sum_{t =1}^{\eta}\ell_t(\bbrtheta^{(1)}) + \min_{\bbrtheta^{(2)} \in \Theta_{B}}\sum_{t = \eta + 1}^{T}\ell_t(\bbrtheta^{(2)})\Bigg\},
\end{align}
where $\ell_t(\bbrtheta)$ is the negative log-likelihood given in \Cref{eq:likelihood}. Then it holds with probability at least $1 - (Tn)^{-2}$ that
\begin{equation}
    |\hat{\eta} - \eta|\leq \frac{Cp_{lb}^{-2}|E|nd_{\max}\log(Tn)}{\lambda_2^2(L_{\mclG})\kappa^2}.
\end{equation}
\enprop

\bprf
The estimator $\hat{\eta}$ is the same as the output of the local refinement algorithm. Under the assumption \eqref{eq:snr_single_cp}, the same arguments in the proof of \Cref{prop:local_refine} can be applied here to show the conclusion.

It should be noted that the estimator $\hat{\eta}$ gives consistent localization because as $T\rightarrow \infty$, we have $\mathcal{B}_T\rightarrow\infty$ and with high probability,
\begin{equation*}
    \frac{|\hat{\eta} - \eta|}{\Delta} \leq \frac{C}{\mathcal{B}_T} = o(1).
\end{equation*}
\eprf

\bnprop[No change point]
\label{prop:0_change}
Suppose we observe $\{(\bfx(t),y_t)\}_{t\in [T]}$ following model \eqref{eq:random_graph} and \eqref{eq:model_bt} and there is no single change point in $[1,T]$. In addition, assume that
\begin{equation}
    T\geq \mathcal{B}_T\frac{|E|nd_{\max}}{p_{lb}^{2}\kappa^2\lambda_2^2(L_{\mclG})}\log (Tn),
    \label{eq:snr_0_cp}
\end{equation}
for a diverging sequence $\{\mathcal{B}_T\}_{T\in \mathbb{Z}_+}$. Then it holds with probability at least $1 - (Tn)^{-2}$ that the DP procedure in \Cref{algo:DP} with tuning parameter $\gamma = C_{\gamma}p_{lb}^{-2}\frac{nd_{\max}}{\lambda_2(L_{\mclG})}\log (Tn)$ will return an empty set.
\enprop

\bprf
Assume that the output $\hat{\mclP}=\{\hat{\eta}_k\}_{k\in [\hat{K}]}$ with $\hat{K}\geq 1$. Let $\mclI_0 = [1, \hat{\eta}_1)$ and $\mclI_{\hat{K}} = [\hat{\eta}_{\hat{K}},T]$. When ${\hat{K}}>1$, let $\mclI_k = [\hat{\eta}_{k-1},\hat{\eta}_{k})$ for $k \in [{\hat{K}} - 1]$. Then by \Cref{lem:l_hat_lower_bt}, with probability at least $1 - (Tn)^{-4}$, we have
\begin{align*}
        \sum_{k = 0}^{{\hat{K}}}L(\hat{\bbrtheta}(\mclI_k),\mclI_k) + {\hat{K}}Cp_{lb}^{-2}\frac{nd_{\max}}{\lambda_2(L_{\mclG})}\log(Tn) &\geq \sum_{k = 0}^{{\hat{K}}}L({\bbrtheta}^*(\mclI_k),\mclI_k) \\
        &= L({\bbrtheta}^*([1,T]),[1,T])\geq L(\hat{\bbrtheta}([1,T]),[1,T]),
\end{align*}
which is contradictory to the definition of $\hat{\mclP}$ as long as $C_{\gamma}>C$.
\eprf

\subsection{Technical lemmas}
\label{sec:lemmas}
This section has three parts:
\begin{enumerate}
    \item \Cref{lem:3case_bt} is a summary of three different cases, and is used in the proof of \Cref{prop:P_K}.
    \item \Cref{sec:excess_risk} contains results on the excess risk of $L(\bbrtheta(\mclI),\mclI)$ in four cases.
    \item \Cref{sec:concentration} contains lemmas on some basic concentration properties related to our problem.
\end{enumerate}

\bnlem
\label{lem:3case_bt}
Given any interval $\mclI = (s,e]\subset [1,T]$ with integers $s,e$ that contains at most two change points. Under all assumptions above, we have 
\begin{enumerate}
    \item If $\mclI$ contains no change points, then with probability at leat $1 - (Tn)^{-2}$ it holds that
    \begin{equation*}
        L(\bbrtheta^*(\mclI),\mclI)\leq L(\hat{\bbrtheta}(\mclI),\mclI) + Cp_{lb}^{-2}\frac{nd_{\max}}{\lambda_2(L_{\mclG})}\log (Tn).
    \end{equation*}
    \item If $\mclI$ contains exactly one change point $\eta_r$ with partition $\mclI_1 = (s,\eta_r]$ and $\mclI_2 = (\eta_r,e]$, then with probability at leat $1 - (Tn)^{-2}$ it holds that
    \begin{equation*}
        L(\bbrtheta^*(\mclI_1),\mclI_1) + L(\bbrtheta^*(\mclI_2),\mclI_2)\leq L(\hat{\bbrtheta}(\mclI),\mclI) + Cp_{lb}^{-2}\frac{nd_{\max}}{\lambda_2(L_{\mclG})}\log (Tn).
    \end{equation*}
    \item If $\mclI$ contains exactly two change points $\eta_{r+1}$ and $\eta_{r+2}$ with partition $\mclI_1 = (s,\eta_{r+1}]$, $\mclI_2 = (\eta_{r+1},\eta_{r+2}]$, and $\mclI_3 = (\eta_{r+2},e]$, then with probability at leat $1 - (Tn)^{-2}$ it holds that
    \begin{equation*}
        \sum_{j = 1}^3L(\bbrtheta^*(\mclI_j),\mclI_j)\leq L(\hat{\bbrtheta}(\mclI),\mclI) + Cp_{lb}^{-2}\frac{nd_{\max}}{\lambda_2(L_{\mclG})}\log (Tn).
    \end{equation*}
\end{enumerate}
\enlem
\bprf
Case 1 is guaranteed by \Cref{lem:l_hat_lower_bt}.

For case 3, since $|\mclI_2|\geq \Delta$, by \Cref{assp:snr_bt}, \Cref{lem:l_hat_lower_bt} and the definition of $\hat{\bbrtheta}$, it holds with probability at least $1 - (Tp)^{-4}$ that
\begin{equation}
    L(\bbrtheta^*(\mclI_2),\mclI_2)\leq L(\hat{\bbrtheta}(\mclI_2),\mclI_2) + C p_{lb}^{-2}\frac{nd_{\max}}{\lambda_2(L_{\mclG})}\log (Tn)\leq L(\hat{\bbrtheta}(\mclI),\mclI_2) + Cp_{lb}^{-2} \frac{nd_{\max}}{\lambda_2(L_{\mclG})}\log (Tn),
\end{equation}
where the second inequality is implied by the definition of $\hat{\bbrtheta}(\mclI_2)$.

For $\mclI_1$, we need to consider two cases. If $|\mclI_1|\geq \frac{C_0|E|}{\lambda_2(L_{\mclG})}\log (Tn)$ where $C_0$ is some fixed absolute constant in the sample size condition in \Cref{assp:glm_bt}, then by \Cref{lem:l_hat_lower_bt}, with probability at least $1 - (Tn)^{-4}$ we have
\begin{equation*}
    L(\bbrtheta^*(\mclI_1),\mclI_1)\leq L(\hat{\bbrtheta}(\mclI_1),\mclI_1) + C p_{lb}^{-2}\frac{nd_{\max}}{\lambda_2(L_{\mclG})}\log (Tn)\leq L(\hat{\bbrtheta}(\mclI),\mclI_1) + C p_{lb}^{-2}\frac{nd_{\max}}{\lambda_2(L_{\mclG})}\log (Tn).
\end{equation*}
Otherwise when $|\mclI_1|< \frac{C_0|E|}{\lambda_2(L_{\mclG})}\log(Tn)$, let $\epsilon_t:=y_t - \frac{\exp(\bfx(t)^\top \bbrtheta^*(t))}{1 + \exp(\bfx(t)^\top \bbrtheta^*(t))}$ and we can get
\begin{align*}
    & L(\bbrtheta^*(\mclI_1),\mclI_1) - \sum_{t\in \mclI_1}\ell_{t}(\hat{\bbrtheta}(\mclI))\\
    =&\sum_{t\in \mclI_1}\ell_{t}(\bbrtheta^*(\mclI_1)) - \sum_{t\in \mclI_1}\ell_{t}(\hat{\bbrtheta}(\mclI))\\
    \leq & \sum_{t\in \mclI_1}\epsilon_t\bfx(t)^\top (\hat{\bbrtheta}(\mclI) - \bbrtheta^*(\mclI_1)) - ce^{-2B}[\bfx(t)^\top (\hat{\bbrtheta}(\mclI) - \bbrtheta^*(\mclI_1))]^2\\
    \leq & \frac{e^{2B}}{4c}\sum_{t\in \mclI_1}[\epsilon_t]^2
    \leq \frac{e^{2B}}{4c}|\mclI_1|\leq C_1 p_{lb}^{-2}\frac{nd_{\max}}{\lambda_2(L_{\mclG})}\log(Tn),
\end{align*}
where the last inequality holds because $|\mclI_1|< \frac{C_0|E|}{\lambda_2(L_{\mclG})}\log(Tn)$ and $|E|\leq nd_{\max}$. Similarly, we can show that
\begin{equation*}
    L(\bbrtheta^*(\mclI_3),\mclI_3) - \sum_{t\in \mclI_3}\ell_{t}(\hat{\bbrtheta}(\mclI))\leq C_1 p_{lb}^{-2}\frac{nd_{\max}}{\lambda_2(L_{\mclG})}\log(Tn).
\end{equation*}
Combining the three facts proves the conclusion for case 3. Similar arguments can be used to prove the conclusion for case 2.
\eprf

\subsubsection{Excess risk}
\label{sec:excess_risk}
\bnlem
\label{lem:convex_bt}
Suppose $\bbrtheta, \bbrtheta(t)^*\in \Theta_{B}$, then
\begin{equation}
    \ell_t(\bbrtheta) - \ell_t(\bbrtheta^*(t))\geq [\frac{\exp(\bfx(t)^\top \bbrtheta^*(t))}{1 + \exp(\bfx(t)^\top \bbrtheta^*(t))} - y_t]\bfx(t)^\top (\bbrtheta - \bbrtheta^*(t)) + ce^{-2B}[\bfx(t)^\top (\bbrtheta - \bbrtheta^*(t))]^2.
\end{equation}
\enlem
\bprf
By Taylor expansion,
\begin{align*}
        &\log(1 + e^{\bfx(t)^\top \bbrtheta}) - \log(1 + e^{\bfx(t)^\top \bbrtheta^*(t)}) \\
        =& [\frac{\exp(\bfx(t)^\top \bbrtheta^*(t))}{1 + \exp(\bfx(t)^\top \bbrtheta^*(t))} - y_t]\bfx(t)^\top (\bbrtheta - \bbrtheta^*(t)) + \frac{\exp(\bfx(t)^\top {\bm \xi})}{[1 + \exp(\bfx(t)^\top {\bm \xi})]^2}[\bfx(t)^\top (\bbrtheta - \bbrtheta^*(t))]^2\\
        \geq &[\frac{\exp(\bfx(t)^\top \bbrtheta^*(t))}{1 + \exp(\bfx(t)^\top \bbrtheta^*(t))} - y_t]\bfx(t)^\top (\bbrtheta - \bbrtheta^*(t)) +  \frac{1}{4e^{2B}}[\bfx(t)^\top (\bbrtheta - \bbrtheta^*(t))]^2.
\end{align*}
where ${\bm \xi}$ is a convex combination of $\bbrtheta$ and $\bbrtheta^*(t)$. Thus, ${\bm \xi}\in \Theta_{B}$ and we also use the facts that $|\bfx(t)^\top \bfv|\leq 2B$ for any $\bfv\in \Theta_B$ and $\frac{e^x}{(1 + e^x)^2}\geq \frac{1}{4e^{|x|}}$.
\eprf

\bnlem
\label{lem:l_hat_lower_bt}
Assume there is no change points in interval $\mclI$, then it holds with probability at least $1 - (Tn)^{-4}$ that
\begin{equation*}
    L(\hat{\bbrtheta}(\mclI),\mclI) - L({\bbrtheta}^*(\mclI),\mclI) = \sum_{t\in \mclI}[\ell_t(\hat{\bbrtheta}) - \ell_t({\bbrtheta}^*)]\geq -Cp_{lb}^{-2}\frac{nd_{\max}}{\lambda_2(L_{\mclG})}\log (Tn),
\end{equation*}
where $C$ is a universal constant that is independent of the choice of $\mclI$.
\enlem
\bprf
Let $\epsilon_t := y_t - \frac{\exp(\bfx(t)^\top \bbrtheta^*(t))}{1 + \exp(\bfx(t)^\top \bbrtheta^*(t))}$. By \Cref{lem:convex_bt}, we have
\begin{equation}
    \begin{split}
     & L({\bbrtheta}^*(\mclI),\mclI) - L(\hat{\bbrtheta},\mclI) \\
    \leq & \sum_{t\in \mclI}\epsilon_t\bfx(t)^\top (\hat{\bbrtheta}(\mclI) - \bbrtheta^*(\mclI)) - ce^{-2B}\sum_{t\in \mclI}[\bfx(t)^\top (\hat{\bbrtheta}(\mclI) - \bbrtheta^*(\mclI))]^2\\
    \leq & \sum_{t\in \mclI}\epsilon_t\bfx(t)^\top (\hat{\bbrtheta}(\mclI) - \bbrtheta^*(\mclI))\\
    \leq & \|\hat{\bbrtheta}(\mclI) - \bbrtheta^*(\mclI)\|_1\max_{i\in [p]}|\sum_{t\in \mclI}\epsilon_tx_i(t)|.
    \end{split}
\end{equation}
When $|\mclI|\gtrsim \frac{C_0|E|}{\lambda_2(L_{\mclG})}\log (Tn)$, by \Cref{lem:estimation_bt}, we have $\|\hat{\bbrtheta}(\mclI) - \bbrtheta^*(\mclI)\|_1\lesssim p_{lb}^{-2}n{\sqrt{\frac{|E|\log(Tn)}{|\mclI|\lambda_2(L_{\mclG})}}}$. Thus, \Cref{lem:epsilon_X_upper_bt} ensures that the first term is upper bounded by $C_1p_{lb}^{-2}n\sqrt{\frac{d_{\max}}{\lambda_2(L_{\mclG})}}\log(Tn)$ where $C_1$ does not depend on $C_0$. Since $\lambda_2(L_{\mclG})\leq 2d_{\max}$, we have
\begin{equation*}
    C_1p_{lb}^{-2}n\sqrt{\frac{d_{\max}}{\lambda_2(L_{\mclG})}}\log(Tn)\leq Cp_{lb}^{-2}\frac{nd_{\max}}{\lambda_2(L_{\mclG})}\log(Tn).
\end{equation*}

When $|\mclI|< \frac{C_0|E|}{\lambda_2(L_{\mclG})}\log (Tn)$, we can bound the difference by
\begin{equation}
    \begin{split}
     & L({\bbrtheta}^*(\mclI),\mclI) - L(\hat{\bbrtheta}(\mclI),\mclI) \\
    \leq & \sum_{t\in \mclI}\epsilon_t\bfx(t)^\top (\hat{\bbrtheta}(\mclI) - \bbrtheta^*(\mclI)) - ce^{-2B}\sum_{t\in \mclI}[\bfx(t)^\top (\hat{\bbrtheta}(\mclI) - \bbrtheta^*(\mclI))]^2\\
    \leq & \frac{e^{2B}}{4c}\sum_{t\in \mclI}\epsilon_t^2
    \leq C_2p_{lb}^{-2}\frac{C_0|E|}{\lambda_2(L_{\mclG})}\log (Tn),
    \end{split}
    \label{eq:appendix_short_interval}
\end{equation}
where we use the fact that $|\epsilon_t|\leq 1$ and $\bbrtheta^*\in \Theta_{B}$ by our assumption, and the basic inequality $ab \leq a^2 + b^2/4$. Since $|E|\leq nd_{\max}$, it holds that
\begin{equation*}
    C_2p_{lb}^{-2}\frac{C_0|E|}{\lambda_2(L_{\mclG})}\log (Tn)\leq Cp_{lb}^{-2}\frac{nd_{\max}}{\lambda_2(L_{\mclG})}\log (Tn).
\end{equation*}

\eprf

\bnlem
\label{lem:no_change}
Under all assumptions in \Cref{thm:consistency_bt}, let $\mclI = (s,e]\subset [1,T]$ be any interval containing no change point. Let $\mclI_1,\mclI_2$ be two intervals such that $\mclI_1\cup \mclI_2 = \mclI$. It holds with probability at least $1 - (Tn)^{-4}$ that
\begin{equation*}
    L(\hat{\bbrtheta}(\mclI_1),\mclI_1) + L(\hat{\bbrtheta}(\mclI_2),\mclI_2) + \gamma \geq L({\bbrtheta}^*(\mclI),\mclI).
\end{equation*}
\enlem
\bprf
If $\mclI<2C_0\frac{|E|\log (Tn)}{\lambda_2(L_{\mclG})}$, following the same arguments in \Cref{lem:l_hat_lower_bt} we have that for $i = 1, 2$, with probability at least $1 - (Tn)^{-4}$,
\begin{equation*}
    L({\bbrtheta}^*(\mclI_i),\mclI_i) - L(\hat{\bbrtheta}(\mclI_i),\mclI_i)\leq Cp_{lb}^{-2}\frac{nd_{\max}}{\lambda_2(L_{\mclG})}\log (Tn).
\end{equation*}
Thus, by the fact that $L({\bbrtheta}^*(\mclI),\mclI) = L({\bbrtheta}^*(\mclI_1),\mclI_1) + L({\bbrtheta}^*(\mclI_2),\mclI_2)$ and $\gamma = C_{\gamma}(K+1)p_{lb}^{-2}\frac{nd_{\max}}{\lambda_2(L_{\mclG})}\log (Tn)$ with $C_{\gamma}$ large enough, the conclusion holds.

Now assume $\mclI>2C_0\frac{|E|\log (Tn)}{\lambda_2(L_{\mclG})}$. We will prove the lemma by contradiction. Assume that
\begin{equation*}
    L(\hat{\bbrtheta}(\mclI_1),\mclI_1) + L(\hat{\bbrtheta}(\mclI_2),\mclI_2) + \gamma < L({\bbrtheta}^*(\mclI),\mclI).
\end{equation*}
By \Cref{lem:convex_bt}, the equation above implies that
\begin{align}
    ce^{-2B}\sum_{t\in \mclI}[\bfx(t)^\top \Delta(t)]^2 < -\gamma + \sum_{t\in \mclI} \epsilon_t \bfx(t)^{\top} \Delta(t) \label{eq:error_terms_2},
\end{align}
where $\epsilon_t := y_t - \frac{\exp(\bfx(t)^\top \bbrtheta^*(t))}{1 + \exp(\bfx(t)^\top \bbrtheta^*(t))}$ and $\Delta_i(t) = \hat{\theta}_i(\mclI) - \theta^*_i(t)$.
For \eqref{eq:error_terms_2}, following the same arguments in the proof of \Cref{lem:1_change}, we can get that with probability at least $1 - (Tn)^{-4}$,
\begin{equation*}
    \begin{split}
        \sum_{t\in \mclI} \epsilon_t \bfx(t)^{\top} \Delta(t)\leq  C\sqrt{\frac{nd_{\max}}{|E|}\log(Tn)}\left[\sum_{t\in \mclI}\|\Delta(t)\|_2^2\right]^{\frac{1}{2}}.
    \end{split}
\end{equation*}
By \Cref{lem:eig_laplacian_gnm}, with probability at least $1 - (Tn)^{-5}$,
\begin{equation*}
    \sum_{t\in \mclI}[\bfx(t)^\top \Delta(t)]^2\geq \frac{c_1\lambda_2(L_{\mclG})}{|E|}\sum_{t\in \mclI}\|\Delta(t)\|_2^2.
\end{equation*}
Thus, let $z = \sum_{t\in \mclI}\|\Delta(t)\|_2^2$ and we have
\begin{equation*}
    \frac{cc_1\lambda_2}{e^{2B}|E|}z + \gamma \leq C\sqrt{\frac{nd_{\max}}{|E|}\log(Tn)} \sqrt{z}\leq \frac{C^2e^{2B}nd_{\max}}{cc_1\lambda_2}\log(Tn) + \frac{cc_1\lambda_2}{4e^{2B}|E|}z,
\end{equation*}
which implies that
\begin{equation*}
    \sum_{t\in \mclI}\|\Delta(t)\|_2^2 + C_1 \frac{p_{lb}^{-1}|E|}{\lambda_2}\gamma \leq C_2 p_{lb}^{-2}\frac{|E|nd_{\max}}{\lambda_2^2}\log(Tn),
\end{equation*}
which is contradictory to the fact that $\gamma = C_{\gamma}p_{lb}^{-2}(K+1)\frac{nd_{\max}}{\lambda_2(L_{\mclG})}\log(Tn)$ for sufficiently large constant $C_{\gamma}$.
\eprf

\bnlem
\label{lem:1_change}
For $\mclI = (s,e)\subset (0,T+1)$, assume that $\mclI$ contains only one change point $\eta$. Denote $\mclI_1 = (s,\eta]$ and $\mclI_2 = (\eta,e]$. Assume that $\|\bbrtheta^*(\mclI_1) - \bbrtheta^*(\mclI_2)\|_2 = \kappa >0$. If
\begin{equation*}
    L(\hat{\bbrtheta}(\mclI),\mclI)\leq L({\bbrtheta}^*(\mclI_1),\mclI_1) + L({\bbrtheta}^*(\mclI_2),\mclI_2) + \gamma,
\end{equation*}
then with probability at least $1 - (Tn)^{-4}$, there exists an absolute constant $C>0$ such that
\begin{equation*}
    \min\{|\mclI_1|,|\mclI_2|\}\leq C\frac{p_{lb}^{-2}|E|}{\kappa^2\lambda_2(L_{\mclG})}[\gamma + \frac{nd_{\max}}{\lambda_2(L_{\mclG})}\log(Tn)].
\end{equation*}
\enlem
\bprf
Without loss of generality, assume $|\mclI_1|\geq |\mclI_2|$. If $|\mclI_2|< C_0\frac{|E|\log (Tn)}{\lambda_2(L_{\mclG})}$ then the conclusion holds automatically, where $C_0$ is the constant in \Cref{lem:estimation_bt} and \Cref{lem:R_cp_bt}, since we can set $C_{\gamma}$ to be sufficiently large (notice that in the worst case, $\kappa^2$ can be as large as $nB^2$). Thus, in what follows we can assume $|\mclI_2|\geq C_0\frac{|E|\log (Tn)}{\lambda_2(L_{\mclG})}$. Let $\epsilon_t := y_t - \frac{\exp(\bfx(t)^\top \bbrtheta^*(t))}{1 + \exp(\bfx(t)^\top \bbrtheta^*(t))}$ and $\Delta_i(t) = \hat{\theta}_i(\mclI) - \theta^*_i(t)$. By the condition of the lemma and \Cref{lem:convex_bt}, we have
\begin{equation*}
    ce^{-2B}\sum_{t\in \mclI}[\bfx(t)^\top \Delta(t)]^2 \leq \gamma + \sum_{t\in \mclI}\sum_{i\in [n]} \epsilon_t x_i(t) \Delta_i(t).
\end{equation*}
\Cref{lem:R_cp_bt} implies that with probability at least $1 - (Tn)^{-4}$, the term on the right hand side satisfies
\begin{equation*}
    \begin{split}
        & \sum_{t\in \mclI}\sum_{i\in [n]} \epsilon_t x_i(t) (\hat{\theta}_i - \theta^*_i(t))\\
        \leq & \sup_{i \in [n]}\left| \frac{\sum_{t\in \mclI}\epsilon_t x_i(t) (\hat{\theta}_i - \theta^*_i(t))}{\sqrt{\sum_{t\in \mclI} (\hat{\theta}_i - \theta^*_i(t))^2}} \right| \sum_{i\in [n]}\sqrt{\sum_{t\in \mclI} (\hat{\theta}_i - \theta^*_i(t))^2}\\
        \leq & C\sqrt{\frac{d_{\max}}{|E|}\log(Tn)} \sum_{i\in [n]}\sqrt{\sum_{t\in \mclI} (\hat{\theta}_i - \theta^*_i(t))^2}
        \leq C\sqrt{\frac{nd_{\max}}{|E|}\log(Tn)}\sqrt{\sum_{t\in \mclI}\|\Delta(t)\|_2^2}.
    \end{split}
\end{equation*}
By \Cref{lem:eig_laplacian_gnm}, $\sum_{t\in \mclI_i}[\bfx(t)^\top \Delta(t)]^2\geq \frac{c_1\lambda_2(L_{\mclG})}{|E|}\sum_{t\in \mclI_i}\|\Delta(t)\|_2^2$ with probability at least $1 - (Tn)^{-5}$ for $i = 1, 2$. Therefore, letting $z = \sum_{t\in \mclI}\|\Delta(t)\|_2^2$, we have
\begin{align*}
    cc_1\frac{\lambda_2(L_{\mclG})}{e^{2B}|E|}z \leq \gamma + c_2\sqrt{\frac{d_{\max}}{|E|}\log(Tn)} \sqrt{z}.
\end{align*}
Solving the inequality above gives 
\begin{equation*}
    \sum_{t\in \mclI}\|\Delta(t)\|_2^2\leq Cp_{lb}^{-2}\frac{|E|}{\lambda_2(L_{\mclG})}[\gamma + \frac{nd_{\max}}{\lambda_2(L_{\mclG})}\log(Tn)],
\end{equation*}
where $C$ is a universal constant that only depends on $c,c_1,c_2$. Since $\sum_{t\in \mclI}\|\Delta(t)\|_2^2\geq \frac{|\mclI_1||\mclI_2|}{|\mclI|}\kappa^2\geq \frac{\kappa^2}{2}|\mclI_2|$, we have $|\mclI_2|\leq \frac{2C}{\kappa^2}p_{lb}^{-2}\frac{|E|}{\lambda_2(L_{\mclG})}[\gamma + \frac{nd_{\max}}{\lambda_2(L_{\mclG})}\log(Tn)]$.
\eprf

\bnlem
\label{lem:2_change}
Under all assumptions in \Cref{thm:consistency_bt}, let $\mclI = (s,e]\subset [1,T]$ be any interval containing exactly two change points $\eta_{r+1}$ and $\eta_{r+2}$, $\mclI_1 = (e,\eta_{r+1}]$, $\mclI_2 = (\eta_{r+1},\eta_{r+2}]$, and $I_3 = (\eta_{r+2},e]$. Let $\kappa_i = \|\bbrtheta^*(\mclI_i) - \bbrtheta^*(\mclI_{i + 1})\|_2$ for $i = 1,2$ and $\kappa = \min\{\kappa_1,\kappa_2\}$. If
\begin{equation*}
    L(\hat{\bbrtheta}(\mclI),\mclI)\leq \sum_{i = 1}^3 L(\bbrtheta^*(\mclI_i),\mclI_i) + 2\gamma,
\end{equation*}
then it holds with probability at least $1 - (Tn)^{-4}$ that
\begin{equation*}
    \max\{|\mclI_1|,|\mclI_3|\}\leq Cp_{lb}^{-2}\frac{|E|}{\lambda_2(L_{\mclG})}[\gamma + \frac{nd_{\max}}{\lambda_2(L_{\mclG})}\log(Tn)].
\end{equation*}
\enlem
\bprf
Without loss of generality, we assume $|\mclI_1|\geq |\mclI_3|$. There are three possible cases: 1. $|\mclI_1|\leq C_0\frac{|E|\log (Tn)}{\lambda_2(L_{\mclG})}$, 2. $|\mclI_3|\geq C_0\frac{|E|\log (Tn)}{\lambda_2(L_{\mclG})}$, and 3. $|\mclI_1|\geq C_0\frac{|E|\log (Tn)}{\lambda_2(L_{\mclG})}\geq |\mclI_3|$ where $C_0$ is the constant in \Cref{lem:estimation_bt} and \Cref{lem:R_cp_bt}. In case 1 the conclusion holds immediately since we can set $C_{\gamma}$ to be large enough. In case 2, the condition in the lemma implies that
\begin{align*}
        ce^{-2B}\sum_{t\in \mclI}[\bfx(t)^\top \Delta(t)]^2 \leq 2\gamma + \sum_{t\in \mclI} \epsilon_t \bfx(t)^{\top} \Delta(t),
\end{align*}
where $\epsilon_t := y_t - \frac{\exp(\bfx(t)^\top \bbrtheta^*(t))}{1 + \exp(\bfx(t)^\top \bbrtheta^*(t))}$ and $\Delta_i(t) = \hat{\theta}_i(\mclI) - \theta^*_i(t)$.

For the term involving $\epsilon_t$, following the same arguments in the proof of \Cref{lem:1_change}, we can get that with probability at least $1 - (Tn)^{-4}$,
\begin{equation*}
    \begin{split}
        \sum_{t\in \mclI} \epsilon_t \bfx(t)^{\top} \Delta(t)
        \leq  C\sqrt{\frac{nd_{\max}}{|E|}\log(Tn)}\left[\sum_{t\in \mclI}\|\Delta(t)\|_2^2\right]^{\frac{1}{2}}.
    \end{split}
\end{equation*}
Let $z = \sum_{t\in \mclI}\|\Delta(t)\|_2^2$. By \Cref{lem:eig_laplacian_gnm}, $\sum_{t\in \mclI}[\bfx(t)^\top \Delta(t)]^2\geq \frac{c_1\lambda_2(L_{\mclG})}{|E|}\sum_{t\in \mclI}\|\Delta(t)\|_2^2$ with probability at least $1 - (Tn)^{-5}$, and thus we have
\begin{equation*}
    \frac{cc_1\lambda_2(L_{\mclG})}{e^{2B}|E|}z\leq C\sqrt{\frac{nd_{\max}}{|E|}\log(Tn)} \sqrt{z} + 2\gamma,
\end{equation*}
which implies that
\begin{equation*}
    \sum_{t\in \mclI}\|\Delta(t)\|_2^2\leq C_1p_{lb}^{-2}\frac{|E|nd_{\max}}{\lambda^2_2(L_{\mclG})}\log(Tn) + C_2\frac{e^{2B}|E|}{\lambda_2(L_{\mclG})}\gamma.
\end{equation*}
Denote $\tilde{\mclI}$ as the shorter one of $|\mclI_1|$ and $|\mclI_2|$. The left hand can be lowered bounded by
\begin{align*}
    \sum_{t\in \mclI}\|\Delta(t)\|_2^2\geq \sum_{t\in \mclI_1\cup \mclI_2}\|\Delta(t)\|_2^2\geq \frac{|\mclI_1||\mclI_2|}{|\mclI_1| + |\mclI_2|}\kappa^2\geq \frac{|\tilde{\mclI}|}{2}\kappa^2.
\end{align*}
If $|\mclI_2|< |\mclI_1|$, then we have
\begin{equation*}
    \frac{|\mclI_2|}{2}\kappa^2\leq C_1p_{lb}^{-2}\frac{|E|nd_{\max}}{\lambda^2_2(L_{\mclG})}\log(Tn) + C_2p_{lb}^{-1}\frac{|E|}{\lambda_2(L_{\mclG})}\gamma,
\end{equation*}
which leads to the bound
\begin{equation*}
    |\mclI_2|\lesssim \frac{p_{lb}^{-2}|E|}{\kappa^2\lambda_2(L_{\mclG})}[\gamma + \frac{nd_{\max}}{\lambda_2(L_{\mclG})}\log(Tn)],
\end{equation*}
and is contradictory to the assumption that $\Delta \geq \mathcal{B}_T p_{lb}^{-4}K\frac{|E|nd_{\max}}{\kappa^2\lambda_2(L_{\mclG})}\log(Tn)$ in \Cref{assp:snr_bt} because of the definition $\gamma = C_{\gamma}p_{lb}^{-2}(K+1)\frac{nd_{\max}}{\lambda_2(L_{\mclG})}\log(Tn)$. Therefore, we have $|\mclI_2|\geq |\mclI_1|$ and by the same arguments,
\begin{equation*}
    |\mclI_1|\lesssim \frac{p_{lb}^{-2}|E|}{\kappa^2\lambda_2(L_{\mclG})}[\gamma + \frac{nd_{\max}}{\lambda_2(L_{\mclG})}\log(Tn)].
\end{equation*}
Since we assume $|\mclI_3|\leq |\mclI_1|$, the desired bound holds.

In case 3, we only need to prove that $|\mclI_1|\leq C\frac{p_{lb}^{-2}|E|}{\kappa^2\lambda_2(L_{\mclG})}[\gamma + \frac{nd_{\max}}{\lambda_2(L_{\mclG})}\log(Tn)]$. Following the same arguments for \Cref{eq:appendix_short_interval}, we can get that with probability at least $1 - (Tn)^{-5}$,
\begin{equation*}
    L(\bbrtheta^*(\mclI_3),\mclI_3)  - L(\hat{\bbrtheta}(\mclI),\mclI_3)\leq  Cp_{lb}^{-2}\frac{nd_{\max}}{\lambda_2(L_{\mclG})}\log(Tn)\leq  \gamma / 3.
\end{equation*}
Therefore, by the condition of the lemma, we have
\begin{equation*}
    L(\hat{\bbrtheta}(\mclI),\mclI_1\cup \mclI_2)\leq \sum_{i = 1}^2 L(\bbrtheta^*(\mclI_i),\mclI_i) + \frac{7}{3}\gamma.
\end{equation*}
Since $\mclI_1\cup \mclI_2$ only contains 1 true change point, the conclusion can be shown by the same arguments of \Cref{lem:1_change}.
\eprf

\bnlem
\label{lem:3more_change}
Under all assumptions in \Cref{thm:consistency_bt}, let $\mclI = (s,e]\subset [1,T]$ be any interval containing $J\geq 3$ change points $\eta_{r+1},\cdots,{r+J}$. Let $\mclI_1 = (e,\eta_{r + 1}]$, $\mclI_j = (\eta_{r + j - 1},\eta_{r + j}]$ for $j = 2,\cdots,J$, and $\mclI_{J+1} = (\eta_{r + J},e]$. Also let $\kappa_j = \|\bbrtheta^*(\mclI_j) - \bbrtheta^*(\mclI_{j + 1})\|_2$ for $j \in [J]$ and $\kappa = \min_{j\in [J]}\{\kappa_j\}$. Then it holds with probability at least $1 - (Tn)^{-4}$ that
\begin{equation*}
    L(\hat{\bbrtheta}(\mclI),\mclI)> \sum_{j = 1}^{J+1} L(\bbrtheta^*(\mclI_j),\mclI_j) + J\gamma,
\end{equation*}
\enlem
\bprf
Without loss of generality, assume that $|\mclI_1|\geq|\mclI_{J+1}|$. Similar to \Cref{lem:2_change}, there are three cases: 1. $|\mclI_1|\leq C_0\frac{|E|\log (Tn)}{\lambda_2(L_{\mclG})}$, 2. $|\mclI_{J+1}|\geq C_0\frac{|E|\log (Tn)}{\lambda_2(L_{\mclG})}$, and 3. $|\mclI_1|\geq C_0\frac{|E|\log (Tn)}{\lambda_2(L_{\mclG})}\geq |\mclI_{J+1}|$ where $C_0$ is the constant in \Cref{lem:estimation_bt} and \Cref{lem:R_cp_bt}. In case 2, we prove the conclusion by contradiction. Assume that
\begin{equation*}
    L(\hat{\bbrtheta}(\mclI),\mclI)\leq \sum_{j = 1}^{J+1} L(\bbrtheta^*(\mclI_j),\mclI_j) + J\gamma
\end{equation*}
We have
\begin{align*}
        ce^{-2B}\sum_{t\in \mclI}[\bfx(t)^\top \Delta(t)]^2 \leq J\gamma + \sum_{t\in \mclI} \epsilon_t \bfx(t)^{\top} \Delta(t),
\end{align*}
where $\epsilon_t := y_t - \frac{\exp(\bfx(t)^\top \bbrtheta^*(t))}{1 + \exp(\bfx(t)^\top \bbrtheta^*(t))}$ and $\Delta_i(t) = \hat{\theta}_i(I) - \theta^*_i(t)$.

For the term that contains $\epsilon_t$, we can bound it as
\begin{align*}
            \sum_{t\in \mclI} \epsilon_t \bfx(t)^{\top} \Delta(t)
        \leq C\sqrt{\frac{nd_{\max}}{|E|}\log(Tn)}\left[\sum_{t\in \mclI}\|\Delta(t)\|_2^2\right]^{\frac{1}{2}},
\end{align*}
with probability at least $1 - (Tn)^{-4}$. Combining the bounds on both terms and use \Cref{lem:eig_laplacian_gnm} lead to a similar inequality in \Cref{lem:2_change} whose solution gives us
\begin{equation*}
    \sum_{t\in \mclI}\|\Delta(t)\|_2^2\leq C_1 p_{lb}^{-2}\frac{|E|nd_{\max}}{\lambda_2^2(L_{\mclG})}\log(Tn) + C_2 J\frac{e^{2B}|E|}{\lambda_2(L_{\mclG})}\gamma.
\end{equation*}
By definition we know that for $1\leq j\leq J$, $|\mclI_j|\geq \Delta$ and thus,
\begin{align*}
    \sum_{t\in \mclI}\|\Delta(t)\|_2^2&\geq \sum_{j= 1}^J\sum_{t\in \mclI_j}\|\Delta(t)\|_2^2\\
    &\geq \sum_{j= 1}^{J - 1}\frac{1}{2}[\sum_{t\in \mclI_j}\|\Delta(t)\|_2^2 + \sum_{t\in \mclI_{j+1}}\|\Delta(t)\|_2^2]\\
    &\geq \sum_{j= 1}^{J - 1}\frac{1}{2}\cdot\frac{|\mclI_j||\mclI_{j + 1}|}{|\mclI_j| + |\mclI_{j + 1}|}\kappa^2 \\
    &\geq \frac{1}{4}(J - 1)\Delta\kappa^2.
\end{align*}
Therefore, we have
\begin{equation*}
    \Delta\kappa^2\leq C_3p_{lb}^{-2}\frac{|E|}{\lambda_2(L_{\mclG})}[\gamma + \frac{nd_{\max}}{J\lambda_2(L_{\mclG})}\log(Tn)].
\end{equation*}
Since we assume $|\mclI_2|\geq |\mclI_3|$, the inequality above contradicts to the assumption that $\Delta\kappa^2 \geq \mathcal{B}_Tp_{lb}^{-4}K\frac{|E|nd_{\max}}{\lambda_2^2(L_{\mclG})}\log(Tn)$ in \Cref{assp:snr_bt}.

In case 1, following the same arguments of \Cref{eq:appendix_short_interval}, we can get that for $j = 1, J+1$, with probability at least $1 - (Tn)^{-5}$,
\begin{equation*}
    L(\bbrtheta^*(\mclI_j),\mclI_j)  - L(\hat{\bbrtheta}(\mclI),\mclI_j)\leq Cp_{lb}^{-2}\frac{C_0|E|}{\lambda_2(L_{\mclG})}\log(Tn)\leq  \gamma / 3.
\end{equation*}
Similar to case 2, we assume that
\begin{equation*}
    L(\hat{\bbrtheta}(\mclI),\mclI)\leq \sum_{j = 1}^{J+1} L(\bbrtheta^*(\mclI_j),\mclI_j) + J\gamma.
\end{equation*}
Therefore,
\begin{equation*}
    \sum_{j = 2}^J L(\hat{\bbrtheta}(\mclI),\mclI_j)\leq \sum_{j = 2}^{J} L(\bbrtheta^*(\mclI_j),\mclI_j) + (J + \frac{2}{3})\gamma.
\end{equation*}
When $J=3$, following same arguments in \Cref{lem:2_change}, we lead to a contradiction that $\Delta\leq Cp_{lb}^{-2}\frac{|E|}{\kappa^2\lambda_2}[\gamma + \frac{nd_{\max}}{\lambda_2}\log(Tn)]$. When $J>3$, we can get the same contradiction by the same arguments for case 2 in this lemma. Case 3 can be handled in a similar manner.
\eprf

\subsubsection{Basic concentrations}
\label{sec:concentration}
First we introduce some results on the empirical risk minimizer of the Bradley-Terry model, which is defined by the constraint MLE
\begin{equation}
    \hat{\bbrtheta} = \argmin_{\theta\in \Theta_{B}} \sum_{i\in [m]}\ell_i(\bbrtheta).
    \label{eq:glm_bt}
\end{equation}

\bnassum 
\label{assp:glm_bt}
Assume that $(\bfx(t),y_t)_{t\in [m]}$ are i.i.d. observations generated from model \eqref{eq:model_bt} with $\bbrtheta^*(t) = \bbrtheta^*\in \Theta_B$ being a constant vector and \eqref{eq:random_graph} and the sample size $m$ satisfies $m\geq C_0 \frac{|E|\log n}{\lambda_2(L_\mclG)}$.
\enassum

Denote $G(\mclG,m)$ as the (weighted) random graph constructed by randomly sampling $m$ edges with replacement from a fixed symmetric, undirected, and binary graph $\mclG([n],E)$ of $n$ nodes.
\bnlem[Laplacian, general graph]
\label{lem:eig_laplacian_g}
Let $A$ be a (weighted) adjacency matrix sampled from the random graph model $G(\mclG,m)$ and $L_{A} = D - A$ be the Laplacian matrix. Denote the eigenvalues of a Laplacian matrix $L$ as $0 = \lambda_1(L) \leq \lambda_2(L)\leq\cdots\leq \lambda_n(L)$ for $L = L_A,L_\mclG$. Suppose $m\geq C_0 \frac{|E|\log n}{\lambda_2(L_\mclG)}$ for some sufficiently large constant $C_0 >0$, then with probability at least $1 - O(n^{-10})$ we have
\begin{equation}
    \frac{m\lambda_2(L_\mclG)}{2|E|}\leq \lambda_2(L_A)\leq \lambda_n(L_A) \leq \frac{3m\lambda_n(L_\mclG)}{|E|}.
\end{equation}
\enlem
\bprf
Consider a partial isometry matrix $R\in \mathbb{R}^{(n-1)\times n}$ that satisfies $RR^\top = I_{n-1}$ and $R\mathbf{1}_n = 0$. By basic algebra we know that ${\rm rank}(R) = n-1$ and $\{R^\top v: v\in \mathbb{R}^{n-1}\} = \{a\mathbf{1}_n:a\in \mathbb{R}\}^{\perp}$. Consider $Y = RL_{\mclG} R^\top$, then the eigenvalues $\{\lambda_i(L_{\mclG})\}_{i=2}^n$ are the same as eigenvalues of $Y$. Since $\mathbb{E}[Y] = \frac{m}{|E|}RL_{\mclG}R^\top$, by matrix Chernoff inequality (e.g., Theorem 5.1.1 in \cite{tropp2015}), we have
\begin{equation}
    \mathbb{P}(\lambda_2(L_A)\leq \frac{m\lambda_2(L_\mclG)}{2|E|}) = \mathbb{P}(\lambda_{\min}(Y)\leq \frac{m\lambda_2(L_\mclG)}{2|E|})\leq n\exp(-\frac{m\lambda_2(L_\mclG)}{8|E|})\leq n^{-10}
    \label{eq:bound_lambda_LA}
\end{equation}
for $m\geq C_0 \frac{|E|\log n}{\lambda_2(L_\mclG)}$ where $C_0 $ is a sufficiently large constant. Similarly, we can show that $\lambda_n(L_A)<\frac{3m\lambda_n(L_\mclG)}{|E|}$ with probability at least $1 - n^{-10}$.
\eprf

\bnlem[Estimation of BTL, general graph]
\label{lem:estimation_bt}
Under \Cref{assp:glm_bt}, for the MLE $\hat{\bbrtheta}$ defined in \Cref{eq:glm_bt}, with probability at least $1 - O(n^{-10})$ we have
\begin{equation}
    \|\hat{\bbrtheta} - \bbrtheta^*\|_2\leq Cp_{lb}^{-2}\sqrt{\frac{n|E|\log n}{m\lambda_2(L_{\mclG})}},\quad \|\hat{\bbrtheta} - \bbrtheta^*\|_1\leq Cp_{lb}^{-2}n\sqrt{\frac{|E|\log n}{m\lambda_2(L_{\mclG})}}.
\end{equation}
\enlem
\bprf
The first inequality is a corollary of Theorem 2 in \cite{shah2015estimationfrompairwisecomps} and \Cref{lem:eig_laplacian_gnm}. Specifically, \cite{shah2015estimationfrompairwisecomps} ensures that with probability at least $1 - O(n^{-12})$,
\begin{equation*}
    \|\hat{\bbrtheta} - \bbrtheta^*\|_2^2\leq Cp_{lb}^{-4}\frac{n\log(n)}{ \lambda_2(L_{A})}.
\end{equation*}
By \Cref{eq:bound_lambda_LA}, $\lambda_2(L_{A})\geq \frac{m\lambda_2(L_\mclG)}{2|E|}$ with probability at least $1 - O(n^{-12})$, so a union bound leads to the conclusion. The second inequality is implied by $\|x\|_1\leq \sqrt{n}\|x\|_2$ for any $x\in \mathbb{R}^n$.
\eprf

As a special case, the random graph model $G(n, m)$ generates random graphs with the vertex set $[n]$ and $m$ edges randomly sampled from the full edge set $E_{full}=\{(i,j):1\leq i<j\leq n\}$. \Cref{lem:eig_laplacian_gnm} gives high probability bounds for the spectra of random graphs following $G(n, m)$.
\bnlem[Laplacian, complete graph]
\label{lem:eig_laplacian_gnm}
Let $A$ be a (weighted) adjacency matrix sampled from the random graph model $G(n, m)$ and $L_A = D - A$ be the Laplacian matrix. Denote the eigenvalues of $L_A$ as $0 = \lambda_1 \leq \lambda_2\leq\cdots\leq \lambda_n$. Suppose $m\geq C_0 n\log n$ for some sufficiently large constant $C_0 >0$, then with probability at least $1 - O(n^{-10})$ we have
\begin{equation}
    \frac{m}{n}\leq \lambda_2(L_A)\leq \lambda_n(L_A) \leq \frac{4m}{n}.
\end{equation}
\enlem
\bprf
Consider a partial isometry matrix $R\in \mathbb{R}^{(n-1)\times n}$ that satisfies $RR^\top = I_{n-1}$ and $R\mathbf{1}_n = 0$. By basic algebra we know that ${\rm rank}(R) = n-1$ and $\{R^\top v: v\in \mathbb{R}^{n-1}\} = \{a\mathbf{1}_n:a\in \mathbb{R}\}^{\perp}$. Consider $Y = RL_AR^\top$, then the eigenvalues $\{\lambda_i\}_{i=2}^n$ are the same as eigenvalues of $Y$. Since $\mathbb{E}[Y] = \frac{2m}{n-1}I_{n-1}$, by matrix Chernoff inequality (e.g., Theorem 5.1.1 in \cite{tropp2015}), we have
\begin{equation}
    \mathbb{P}(\lambda_2(L_A)\leq \frac{m}{n-1}) = \mathbb{P}(\lambda_{\min}(Y)\leq \frac{m}{n-1})\leq (n-1)e^{-\frac{m}{8(n-1)}}\leq n^{-10}
    \label{eq:bound_lambda_LA_complete_graph}
\end{equation}
for $m\geq C_0 n\log n$ where $C_0 $ is a sufficiently large constant. Similarly, we can show that $\lambda_n(L_A)\leq 4m/n$ with probability at least $1 - n^{-10}$.
\eprf

\bnlem[Estimation of BTL, complete graph]
\label{lem:estimation_bt_complete_graph}
Under \Cref{assp:glm_bt}, for the MLE $\hat{\bbrtheta}$ defined in \Cref{eq:glm_bt}, with probability at least $1 - O(n^{-10})$ we have
\begin{equation}
    \|\hat{\bbrtheta} - \bbrtheta^*\|_2\leq Cp_{lb}^{-2}n\sqrt{\frac{\log n}{m}},\quad \|\hat{\bbrtheta} - \bbrtheta^*\|_1\leq Cp_{lb}^{-2}n^{3/2}\sqrt{\frac{\log n}{m}}.
\end{equation}
\enlem
\bprf
The first inequality is a corollary of Theorem 2 in \cite{shah2015estimationfrompairwisecomps} and \Cref{lem:eig_laplacian_gnm}. Specifically, \cite{shah2015estimationfrompairwisecomps} ensures that with probability at least $1 - O(n^{-12})$,
\begin{equation*}
    \|\hat{\bbrtheta} - \bbrtheta^*\|_2^2\leq Cp_{lb}^{-4}\frac{n\log(n)}{ \lambda_2(L_{A})}.
\end{equation*}
By \Cref{eq:bound_lambda_LA}, $\lambda_2(L_{A})\geq m/n$ with probability at least $1 - O(n^{-12})$, so a union bound leads to the conclusion. The second inequality is implied by $\|x\|_1\leq \sqrt{n}\|x\|_2$ for any $x\in \mathbb{R}^n$.
\eprf

In what follows, we prove some concentration properties related to $\epsilon_t := y_t - \frac{\exp(\bfx(t)^\top \bbrtheta^*(t))}{1 + \exp(\bfx(t)^\top \bbrtheta^*(t))}$.

\bnlem
\label{lem:R_cp_bt}
Under all assumptions in \Cref{thm:consistency_bt}, let $\mclI = [1,T]$ be an integer interval such that $|\mclI|\geq c_0(R+1) \frac{|E|}{\lambda_2(L_{\mclG})}\log(Tn)$ and $R$ be a fixed integer. Denote $S_{\mclI,R} = \{\bfv\in \mathbb{R}^{|\mclI|}:\|\bfv\|_2=1,\ \|D\bfv\|_0 = R,\ \min\{k: v_j\neq v_{j + k}\}\geq c_0 \frac{|E|}{\lambda_2(L_{\mclG})}\log(Tn)\}$ and $\epsilon_t := y_t - \frac{\exp(\bfx(t)^\top \bbrtheta^*(t))}{1 + \exp(\bfx(t)^\top \bbrtheta^*(t))}$. Then for some sufficiently large constant $C$, it holds with probability at least $1 - (Tn)^{-2R-10}$ that
\begin{equation*}
    \max_{i\in [p]}\sup_{\bfv\in S_{\mclI, R}}\sum_{t\in I}v_t\epsilon_tx_i(t)\leq C\sqrt{\frac{d_{\max}R\log (Tn) }{|E|}}.
\end{equation*}
\enlem
\bprf
Since $\|D\bfv\|_0 = R$, $\{v_t\}$ is piece-wise constant over $\mclI = [1,T]$ and has $R$ change points that have at most $\binom{T}{R}$ possible choices of locations. Let $\{\eta_k\}_{k\in [R]}$ be the change points of $\{v_t\}$ and $\mclS(\{\eta_k\}_{k\in [R]})$ the linear subspace of $\mclR^{|\mclI|}$ that contains all piecewise-linear sequences over $\mclI$ whose change points are $\{\eta_k\}_{k\in [R]}$. Let $\mclN_{\delta}(\{\eta_k\}_{k\in [R]})$ be a $\delta$-net of $\mclS(\{\eta_k\}_{k\in [R]})\cap \mclS^{|\mclI|}$ where $\mclS^{|\mclI|}$ is the unit sphere in $\mclR^{|\mclI|}$. By Lemma 4.1 in \cite{Pollard1990EmpiricalPT}, since $\mclS(\{\eta_k\}_{k\in [R]})$ is an affine space with dimension $R+1$, we can pick a $\delta$-net $\mclN_{\delta}(\{\eta_k\}_{k\in [R]})$ such that $|\mclN_{\delta}(\{\eta_k\}_{k\in [R]})|\leq (\frac{3}{\delta})^{R+1}$.

Taking $\delta= \frac{1}{|\mclI|}$, then for any fixed $i\in [n]$ and fixed set of change points $\{\eta_k\}_{k\in [R]}$, we have
\begin{align*}
    &\mathbb{P}\left[ \sup_{\bfv\in \mclS_{\mclI, R}} \sum_{t\in \mclI}v_t\epsilon_tx_i(t)\geq C\sqrt{d_{\max}R\log (Tn)/|E|} \right]\\
    \leq & \mathbb{P}\left[ \sup_{\bfu\in \mclN_{1/|\mclI|}(\{\eta_k\}_{k\in [R]})} |\sum_{t\in \mclI}u_t\epsilon_tx_i(t)| + \sup_{\bfv\in \mclS_{\mclI, R}}\inf_{\bfu\in \mclN_{1/|\mclI|}}|\sum_{t\in \mclI}(v_t - u_t)\epsilon_tx_i(t)| \geq C\sqrt{d_{\max}R\log (Tn)/|E|} \right]\\
    \leq & \mathbb{P}\left[ \sup_{\bfu\in \mclN_{1/|\mclI|}(\{\eta_k\}_{k\in [R]})} |\sum_{t\in \mclI}u_t\epsilon_tx_i(t)| + \sup_{\bfv}\inf_{\bfu}\|\bfv - \bfu\|_1\max_{t\in \mclI}|\epsilon_tx_i(t)| \geq C\sqrt{d_{\max}R\log (Tn)/|E|} \right]\\
    \leq & \mathbb{P}\left[ \sup_{\bfu\in \mclN_{1/|\mclI|}(\{\eta_k\}_{k\in [R]})} |\sum_{t\in \mclI}u_t\epsilon_tx_i(t)| + \frac{\sqrt{|\mclI|}}{|\mclI|}\cdot\max_{t\in \mclI}|\epsilon_tx_i(t)| \geq C\sqrt{d_{\max}R\log (Tn)/|E|} \right]\\
    \leq & \mathbb{P}\left[ \sup_{\bfu\in \mclN_{1/|\mclI|}(\{\eta_k\}_{k\in [R]})} |\sum_{t\in \mclI}u_t\epsilon_tx_i(t)| \geq C\sqrt{d_{\max}R\log (Tn)/|E|} \right] \\
    &\quad \times \mathbb{P}\left[ \max_{t\in \mclI} |\epsilon_tx_i(t)| < C\sqrt{d_{\max}R|\mclI|\log (Tn)/|E|}\right] + \mathbb{P}\left[ \max_{t\in \mclI} |\epsilon_tx_i(t)| \geq C\sqrt{d_{\max}R|\mclI|\log (Tn)/|E|}\right]
\end{align*}
Since $\|\bfx(t)\|_{\infty}\leq 1$ and $|\epsilon_t|\leq 2$ under Model \eqref{eq:model_bt}, we can make $C$ sufficiently large so that $\mathbb{P}\left[ \max_{t\in \mclI} |\epsilon_tx_i(t)| \geq C\sqrt{R|\mclI|\log (Tn)/n}\right] = 0$. Therefore,
\begin{align*}
    &\mathbb{P}\left[ \sup_{\bfv\in \mclS_{\mclI, R}} \sum_{t\in \mclI}v_t\epsilon_tx_i(t)\geq C\sqrt{d_{\max}R\log (Tn)/|E|} \right]\\
    \leq & (3|\mclI|)^{R+1}\sup_{\bfu\in \mclN_{1/|\mclI|}(\{\eta_k\}_{k\in [R]})}\mathbb{P}\left[|\sum_{t\in \mclI}u_t\epsilon_tx_i(t)| \geq C\sqrt{d_{\max}R\log (Tn)/|E|} \right] \\
    \leq & (3|\mclI|)^{R+1}\times \max\{2\exp\left[ -\frac{C R\log(Tn)}{\sum_{t\in \mclI}u_t^2} \right], (Tn)^{-3R-12}\}\\
    \leq & C_2\exp(-C_2 R\log (Tn) + R\log (3|\mclI|)),
\end{align*}
where in the second inequality we use \Cref{lem:fixed_v_deviation_bt}. Therefore, for the given interval $\mclI\subset [1,T]$, it holds that
\begin{equation*}
    \mathbb{P}(\mclB_R(\mclI))\leq {\binom{T}{R}} C_2\exp(-C_3 R\log (Tn))\leq (Tn)^{-2R-10},
\end{equation*}
where the event $\mclB_R(\mclI)):=\{\max_{i\in [n]}\sup_{v\in \mclS_{\mclI, R}}\sum_{t\in \mclI}v_t\epsilon_tx_i(t)\geq C\sqrt{d_{\max}R\log (Tn)/|E|}\}$ for some sufficiently large universal constant $C$.
\eprf

\bnlem
\label{lem:fixed_v_deviation_bt}
Let $\epsilon_t = y_t - \frac{\exp(\bfx(t)^\top \bbrtheta^*(t))}{1 + \exp(\bfx(t)^\top \bbrtheta^*(t))}$. Under all assumptions in \Cref{thm:consistency_bt}, for any fixed integer interval $\mclI\subset [1,T]$ such that $|\mclI|\geq c_0 (R+1) \frac{|E|}{\lambda_2(L_{\mclG})}\log(Tn)$ for some sufficiently large constant $c_0>0$ and any fixed $\bfv\in D_{\mclI, R}$ where $D_{\mclI,R} = \{\bfv\in \mathbb{R}^{|\mclI|}:\|D \bfv\|_0 = R,\ \min\{k: v_j\neq v_{j + k}\}\geq c_0 \frac{|E|}{\lambda_2(L_{\mclG})}\log(Tn)\}$ with a fixed integer $R$, it holds for any $\kappa>0$ that
\begin{equation*}
    \max_{i\in [n]}\mathbb{P}\left[|\sum_{t\in \mclI}v_t\epsilon_tx_i(t)|\geq \kappa\right]\leq \max\{2\exp(-\frac{C|E|\kappa^2}{d_{\max}\sum_{t\in \mclI}v_t^2}), (Tn)^{-3R-12}\}.
\end{equation*}
\enlem
\bprf
Following the same arguments in the proof of \Cref{lem:epsilon_X_upper_bt}, we have index set of nonzero terms $\mclI_i$ for each $i\in [n]$. Furthermore, let $\{\mclJ_k\}_{k\in [R+1]}$ be the $R+1$ subintervals such that for each $k$, $v_j$ takes identical values for all $j\in \mclJ_k$. Since $R$ is fixed, by similar arguments we can prove that uniformly for $k\in [R+1]$ and $i\in [n]$, we have $|\mclI_i\cap \mclJ_k|\leq \frac{3d_{\max}}{|E|}|\mclJ_k|$ with probability at least $1 - (Tn)^{-4R-13}$. Now we condition on this event.

By definition, $\mathbb{E}[\epsilon_t|\bfx(t)] = 0$, so for each $i\in [n]$, if we let $S_i(t) = \sum_{j\in [t]} v_{l_{i,t}}\epsilon(l_{i,t})x_i(l_{i,t})$ for $t\in[|\mclI_i|]$ and $S_i(0) =0$, then $\{S_i(t)\}$ is a martingale with respect to the filtration $\{\mclF_t:\mclF_t = \sigma(\bfx(l_{i,1}),\cdots, \bfx(l_{i,t}))\}$. Furthermore, for any $t\in [1,T]$,
\begin{equation*}
    |S_i(t) -S_i(t-1)| \leq |v_{l_{i,t}} x_i(l_{i,t})| \leq |v_{l_{i,t}}|.
\end{equation*}
Thus by \Cref{lem:azuma} we have
\begin{equation*}
     \mathbb{P}\left[|\sum_{t\in \mclI}v_t\epsilon_tx_i(t)|\geq \kappa\right]\leq 2\exp(-\frac{C\kappa^2}{\sum_{t\in \mclI_i}v_t^2}).
\end{equation*}
Now by the fact that $|\mclI_i\cap \mclJ_k|\leq \frac{3d_{\max}}{|E|}|\mclJ_k|$ for each $i,k$, we have $\sum_{t\in \mclI_i}v_t^2\leq \frac{3d_{\max}}{|E|}\sum_{t\in \mclI}v_t^2$. Then the conclusion follows from a union bound.
\eprf

\bnlem[General graph]
\label{lem:epsilon_X_upper_bt}
Let $\epsilon_t = y_t - \frac{\exp(\bfx(t)^\top \bbrtheta^*(t))}{1 + \exp(\bfx(t)^\top \bbrtheta^*(t))}$. Under all assumptions above, for any integer interval $\mclI\subset [1,T]$ such that $|\mclI|\geq C_0\frac{|E|}{\lambda_2(L_{\mclG})}\log(Tn)$ for some sufficiently large constant $C_0>0$, it holds with probability at least $1 - (Tn)^{-10}$
\begin{equation*}
    \max_{i\in [n]}|\sum_{t\in \mclI}\epsilon_tx_i(t)|\leq \sqrt{\frac{d_{\max}}{|E|}|\mclI|\log(Tn)}.
\end{equation*}
\enlem
\bprf
By the assumptions above, i.e., in the comparison graph at each time point a single edge is uniformly randomly picked from the edge set $E$ of $\mclG$, we know that $\mathbb{P}[|x_i(t)|=1] = \frac{d_i}{|E|}\leq \frac{d_{\max}}{|E|}$. Therefore, it follows from a Chernoff inequality (\Cref{lem:chernoff}) that for each $i\in [n]$ with probability at least $1 - (Tn)^{-12}$,
\begin{equation*}
    \sum_{t\in \mclI}|x_i(t)| - \frac{d_i}{|E|}|\mclI|\leq c\sqrt{\log(Tn)}\cdot \sqrt{\frac{d_i}{|E|}|\mclI|}.
\end{equation*}
Since $\lambda_2(L_{\mclG})\leq 2d_{\max}$ and $|\mclI|\geq C_0\frac{|E|}{\lambda_2(L_{\mclG})}\log(Tn)$, we have $|\mclI|\geq c_0\frac{|E|}{d_{\max}}\log(Tn)\geq c_0\frac{|E|d_i}{d^2_{\max}}\log(Tn)$ and thus
\begin{equation*}
    \sum_{t\in \mclI}|x_i(t)| - \frac{d_i}{|E|}|\mclI|\leq c\sqrt{\log(Tn)}\cdot \sqrt{\frac{d_i}{|E|}|\mclI|}\leq C\frac{d_{\max}}{|E|}|\mclI|,
\end{equation*}
which implies that with probaility at least $1 - (Tn)^{-11}$, it holds uniformly for all $i\in [n]$ that in summation $\sum_{t\in \mclI}\epsilon_tx_i(t)$ there are at most $C|\mclI|\frac{d_{\max}}{|E|}$ nonzero terms. 

Now we condition on this event and denote for each $i\in [n]$ the index set of nonzero terms as $\mclI_i$. Thus we have $\sum_{t\in \mclI}\epsilon_tx_i(t) = \sum_{t\in \mclI_i}\epsilon_tx_i(t)$. For each $\mclI_i$, we write its elements as $l_{i,t}$ for $t\in [|\mclI_i|]$ such that $l_{i,1}<l_{i,1}<\cdots<l_{i,|\mclI_i|}$.

By definition, $\mathbb{E}[\epsilon_t|\bfx(t)] = 0$, so for each $i\in [n]$, if we let $S_i(t) = \sum_{j\in[t]} \epsilon(l_{i,t})x_i(l_{i,t})$ for $t\in[|\mclI_i|]$ and $S_i(0) =0$, then $\{S_i(t)\}$ is a martingale with respect to the filtration $\{\mclF_t:\mclF_t = \sigma(\bfx(l_{i,1}),\cdots, \bfx(l_{i,t}))\}$. Furthermore, for any $t\in [1,T]$,
\begin{equation*}
    |S_i(t) -S_i(t-1)| \leq |x_i(l_{i,t})| \leq 1.
\end{equation*}
Thus by Azuma's inequality (\Cref{lem:azuma}) and a union bound we can get the conclusion.
\eprf

\bnlem
\label{lem:chernoff}
Suppose $Z_1,\cdots,Z_s$ are independent random variables with zero expectation and variance $\mathbb{E}Z_i^2 = \sigma^2_i$ satisfying $|Z_i|\leq 1$ almost surely, then
\begin{equation*}
    \mathbb{P}\{| \sum_{i \in [s]}Z_i |\geq u\sigma\} \leq C\max\{e^{-cu^2},e^{-cu\sigma}\},
\end{equation*}
where $\sigma^2 = \sum_{i\in [s]}\sigma_i^2$, and $C,c>0$ are universal constants. In particular, for $u\leq \sigma$, we have
\begin{equation*}
    \mathbb{P}\{| \sum_{i \in [s]}Z_i |\geq u\sigma\} \leq Ce^{-cu^2}.
\end{equation*}
\enlem
\bprf
See Theorem 2.1.3 in \cite{tao2012}.
\eprf

A sequence of random variables $\{D_k\}_{k\in \mathbb{Z}_+}$ is called a martingale difference if there exists a martingale $(Z_k,\mclF_k)_{k\in \mathbb{Z}_+}$ such that $D_k = Z_k - Z_{k-1}$. The following result is well-known in high-dimensional statistics \citep{wainwright2019book}. We include the proof for completeness and the convenience of readers.

\bnlem[Azuma's Inequality or Azuma-Hoeffding Inequality]
\label{lem:azuma}
Suppose $\{D_k\}_{k\in \mathbb{Z}_+}$ is a martingale difference. If $D_k\in (a_k,b_k)$ almost surely for some $a_k<b_k$, then
\[
\mathbb{P}\left( \left|\sum_{k=1}^n D_k\right|\geq t \right) \leq 2\exp\left\lbrace -\frac{2t^2}{\sum_k(b_k-a_k)^2} \right\rbrace.
\]
\enlem
\bprf
$D_k\in (a_k,b_k)$ almost surely implies that for almost all $\omega\in \Omega$, the conditional variable $(D_k|\mclF_{k-1})(\omega)\in (a_k,b_k)$ almost surely, where $(D_k|\mclF_{k-1})(\omega)$ is defined using regular conditional distributions. By the Hoeffding's bound, $(D_k|\mclF_{k-1})(\omega)$ is sub-Gaussian with parameter $\sigma^2 = (b_k-a_k)^2/4$, for almost all $\omega$. Therefore by the definition of sub-Gaussian random variables, we have that for almost all $\omega$,
\[
\mathbb{E}\left[ \exp\{\lambda (D_k|\mclF_{k-1})(\omega)\} \right]\leq \exp\left\lbrace \lambda^2\frac{(b_k-a_k)^2}{8} \right\rbrace.
\]
By the property of regular conditional distributions,
\[
\mathbb{E}\left[e^{\lambda D_k}|\mclF_{k-1}\right](\omega) = \mathbb{E}\left[ \exp\{\lambda (D_k|\mclF_{k-1})(\omega)\} \right], \text{almost surely}.
\]
Therefore
\[
\mathbb{E}\left[e^{\lambda D_k}|\mclF_{k-1}\right] \leq \exp\left\lbrace \lambda^2\frac{(b_k-a_k)^2}{8} \right\rbrace, \text{almost surely}.
\]
Now let $\nu_k^2 = (b_k-a_k)^2/4$ and $\alpha_k = 0$ in Theorem \ref{lem:bdd_ineq} and we can prove the inequality.
\eprf

A random variable $X$ with $\mathbb{E}=\mu$ is called sub-exponential with parameters $\nu^2$ and $\alpha$, or ${\rm SE}(\nu^2,\alpha)$ for brevity, if
\begin{equation*}
    \mathbb{E}[e^{\lambda (X - \mu)}]\leq e^{\lambda^2\nu^2 / 2},\ \forall |\lambda|\leq \frac{1}{\alpha}.
\end{equation*}

\bnlem
\label{lem:bdd_ineq}
Let $\{(D_k,\mclF_k),k\in \mathbb{Z}_+\}$ be a martingale difference s.t.
\begin{equation}
\mathbb{E}\left[ e^{\lambda D_k}|\mclF_{k-1} \right]\leq e^{\lambda^2\nu_k^2/2},\ \forall |\lambda|\leq \frac{1}{\alpha_k},
\label{eq:cond}
\end{equation}
almost surely. Then
\begin{enumerate}
\item[1)] $\sum_{k=1}^n D_k\in {\rm SE}(\sum_k \nu_k^2,\max_k \alpha_k)$;
\item[2)] \[
\mathbb{P}(|\sum_k D_k|\geq t)\leq \begin{cases}
2\exp\left\lbrace -\frac{t^2}{2\sum_k \nu_k^2} \right\rbrace,\ t\leq \frac{\sum_k \nu_k^2}{\max_k \alpha_k},\\
2\exp\left\lbrace -\frac{t}{2\max_k \alpha_k} \right\rbrace,\ t> \frac{\sum_k \nu_k^2}{\max_k \alpha_k}.
\end{cases}
\]
\end{enumerate}
\enlem
\bprf
1). By the iterated law of expectation
\begin{align*}
\mathbb{E}\left[e^{\lambda\sum_{k=1}^nD_k}\right] &= \mathbb{E}\left[ \mathbb{E}\left[ e^{\lambda\sum_{k=1}^nD_k}|\mclF_{n-1} \right] \right]\\
& = \mathbb{E}\left[ \exp\{\lambda\sum_{k=1}^{n-1}D_k\}\mathbb{E}\left[e^{\lambda D_n} |\mclF_{n-1} \right] \right]\\
& \leq \mathbb{E}\left[ \exp\{\lambda\sum_{k=1}^{n-1}D_k\}e^{\lambda^2\nu_n^2/2} \right]\\
& = e^{\lambda^2\nu_n^2/2}\mathbb{E}\left[e^{\lambda\sum_{k=1}^{n-1}D_k}\right],\ {\rm for} |\lambda| <\frac{1}{\alpha_n},
\end{align*}
where we use the fact that $\exp\{\lambda\sum_{k=1}^{n-1}D_k\}\in \mclF_{n-1}$ and \eqref{eq:cond}. Repeating the same procedure for $k = n-1,\cdots,2$, we can get
\[
\mathbb{E}\left[e^{\lambda\sum_{k=1}^nD_k}\right] \leq e^{\lambda^2\frac{\sum_{k=1}^n\nu_k^2}{2}}, \ {\rm for} |\lambda| <\frac{1}{\max_k\alpha_k}.
\]
2) Use the property of sub-exponential random variables and 1).
\eprf



\end{document}